# Lexical Access Model for Italian

## Modeling human speech processing: identification of words in running speech toward lexical access based on the detection of landmarks and other acoustic cues to features

Maria-Gabriella Di Benedetto[1,2], Stefanie Shattuck-Hufnagel[1], Jeung-Yoon Choi[1], Luca De Nardis[2], Javier Arango[3], Ian Chan[3], and Alec DeCaprio[3]

[1] Massachusetts Institute of Technology (MIT), Cambridge, MA, United States

[2] DIET Department, Sapienza University of Rome, Rome, Italy

[3] Radcliffe Institute for Advanced Study, Harvard University, Cambridge, MA, United States

### Author Note

Maria-Gabriella Di Benedetto https://orcid.org/0000-0003-1523-5083

Stefanie Shattuck-Hufnagel https://orcid.org/0000-0003-0991-5541

Luca De Nardis https://orcid.org/0000-0001-9286-8744

We have no known conflict of interest to disclose.

**Correspondence concerning this artide should be addressed to** Maria-Gabriella Di Benedetto, DIET Department, Sapienza University of Rome, Via Eudossiana 18, 00184, Rome, Italy. E-mail: mariagabriella.dibenedetto@uniroma1.it

Modelling the process that a listener actuates in deriving the words intended by a speaker requires setting a hypothesis on how lexical items are stored in memory. This work aims at developing a system that imitates humans when identifying words in running speech and, in this way, provide a framework to better understand human speech processing. We build a speech recognizer for Italian based on the principles of Stevens' model of Lexical Access in which words are stored as hierarchical arrangements of distinctive features (Stevens, K. N. (2002). "Toward a model for lexical access based on acoustic landmarks and distinctive features," J. Acoust. Soc. Am., 111(4):1872–1891). Over the past few decades, the Speech Communication Group at the Massachusetts Institute of Technology (MIT) developed a speech recognition system for English based on this approach. Italian will be the first language beyond English to be explored; the extension to another language provides the opportunity to test the hypothesis that words are represented in memory as a set of hierarchically-arranged distinctive features, and reveal which of the underlying mechanisms may have a language-independent nature. This paper also introduces a new Lexical Access corpus, the LaMIT database, created and labeled specifically for this work, that will be provided freely to the speech research community. Future developments will test the hypothesis that specific acoustic discontinuities – called *landmarks* – that serve as cues to features, are language independent, while other cues may be language-dependent, with powerful implications for understanding how the human brain recognizes speech.

## I. Introduction

Understanding how humans process speech and, in particular, how listeners locate and identify words in running speech, is challenging, because the surface phonetic form of a word varies from one utterance to another. This variability arises from many different factors, including speaker-related characteristics, styles and emotions, noise, word context in the sentence, and others (Burki, 2018). Modeling these transformations in the acoustic properties of words and their timing is still an unmet challenge, even after many decades of strenuous research investigation in the field. Ironically, the relatively recent remarkable progress made in the field of automatic speech recognition by machines, that has led to the development of user-friendly products and applications available to all, has generated the illusional belief that technology has disentangled the mystery around language in both the cognitive systems that process it and the human brain mechanisms that serve these processes. Automatic speech recognition systems show steady improvements in recognition performance; The methods they adopt are based, roughly speaking, on statistical analyses and training on large databases, and take great benefit from the development of Machine Learning, Deep Learning, Deep Neural Nets, and advances in the field of data science addressing the analysis of massive datasets. As such, these systems take little advantage – or no advantage at all – of the knowledge gathered over many decades of research focused on theoretical development of phonology, articulatory phonetics and acoustic phonetics, as well as of extensive results obtained by experimental testing, and, critically, make no attempt to model human speech processing. As a result, this tangible advancement has not gone hand-in-hand with a progress in explaining how we human listeners manage to perform so well, even in the most challenging listening situations. The strong shift of focus and resources toward solutions-driven speech research, based largely on statistical analysis of very large speech corpora, has greatly benefitted the development of the technology. Regrettably, however, this has seriously slowed the expansion of our understanding of all that data as evidence for models of human speech processing. As a result, the data have not been applied to the task of explaining how listeners recover from the signal the information that is needed to perform the seemingly simple task of identifying one word vs. another. Despite the development of the technology and the results obtained, this aspect of spoken language is still veiled in mystery. Cognitive systems that process language and the human brain mechanisms that serve these processes are still poorly understood, and we are still far from a quantitative theory that explains the modifications of the acoustic properties of a word when pronounced in different contexts in running speech. The predicted acoustic properties of a word may often be unavailable in the speech signal, due to the acoustic modifications that the word undergoes and that shift it from a "citation form"[1].

Consider the simple example of a word containing a vowel. As is well known, the acoustic parameters that are used nowadays to characterize vowels can vary substantially from one speaker to another, and in particular among speakers with long vocal tracts vs. speakers with shorter vocals tracts. As a consequence, a vowel may only show part - or none- of the expected canonical properties, and may even end up resembling another vowel. A word may then be missed, or misinterpreted as another word, if two competing words, forming a minimal pair that contrast only by virtue of the distinction between these two vowels, are present in the lexicon. In the absence of a theoretical understanding of how systematic variability arises, the statistical approach provides a solution.

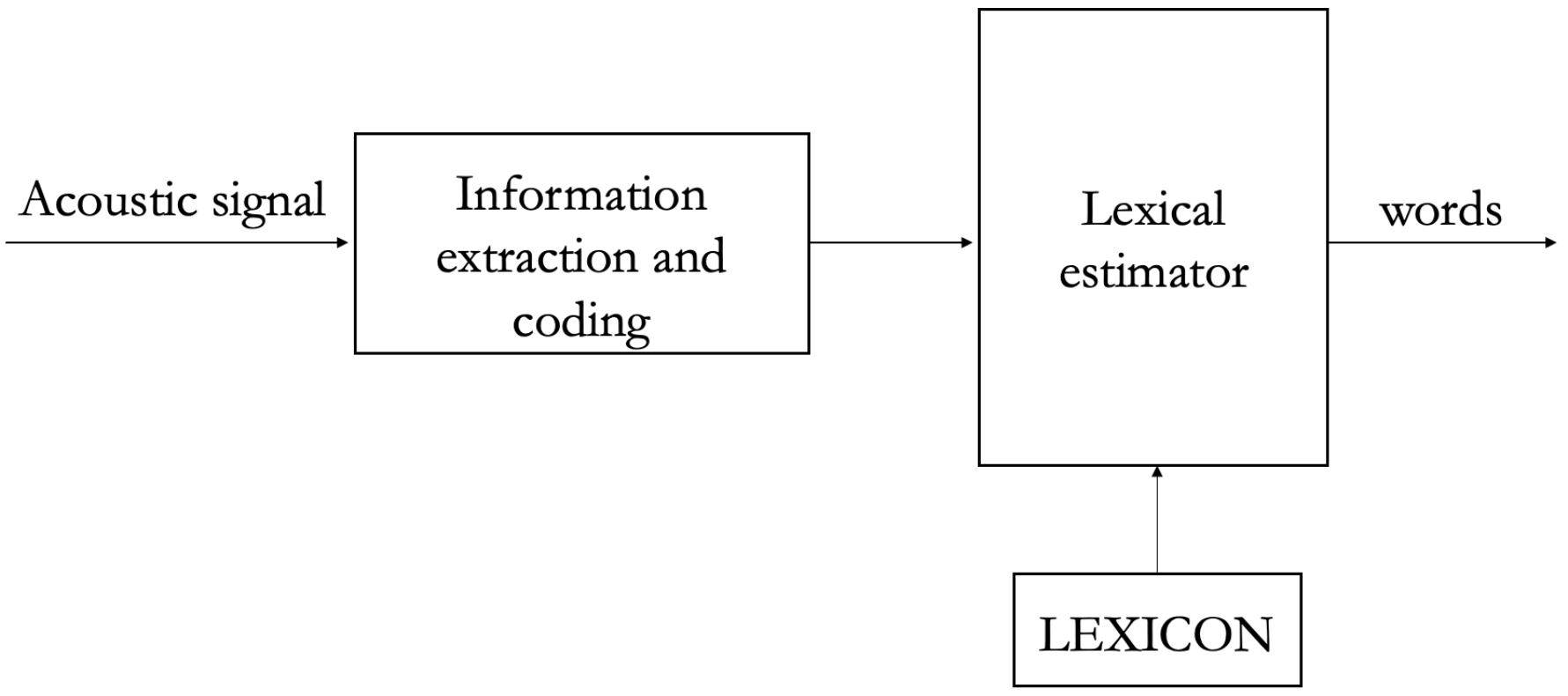

Figure 1 – Model of Lexical Access

The proposed framework focuses on developing a system that mimics the process by which human listeners identify words in running speech, and in this way provides tools to better understand human speech processing and representation. Modeling the process by which listeners formulate lexical hypotheses, based on the analysis of an acoustic speech signal input, is in fact a crucial component of the entire human speech perception process and of the lexical access process in particular. A model for lexical access, i.e. the systematic specification of lexical hypotheses, can be pictured as composed of three fundamental blocks: a lexicon of words for a given language, a processor that extracts and codes the information from an incoming acoustic signal, and an estimator to link the latter with the former (see Fig. 1). In order to retrieve a word, the form in which the information is extracted from the acoustic signal must match the form in which lexical information is stored in the lexicon, i.e. in the memory of the speaker. We address the problem of human lexical access modeling, that is, the design of the three above systems that must communicate with each other, under the peculiar requirements characterizing human speech perception. These include e.g. its robustness to acoustic noise, to dialects, to individual speaker variations, to speaker-specific speaking styles, and to modifications typically introduced by the speaker in the acoustic pattern of a word in running speech, in particular at word boundaries, due to coarticulation effects by which neighboring sounds influence one another, as well as its location in the prosodic structure. Last but not least, the design should be language-independent; there is general agreement that the process that a listener actuates in deriving words intended by a speaker is universal, even though, at the same time, specific properties by which a speaker encodes lexical items, and that are extracted from the speech signal by a listener, may be in part language-specific. In other words, the way the human brain functions when accessing the lexicon of a certain language does not depend on the language itself, although, as we all know from experience, different languages are characterized by different properties and sounds, and, therefore, the information stored in memory, as well as the retrieved information from the acoustic signal, must reflect the specific properties of that particular language.

[1] In this paper we use the term 'citation' to refer to the way a word or sound is produced in a particularly clear context, such as when a consonant or glide is produced in an intervocalic context before a stressed vowel, or when a vowel is produced in prominent position in the intonational phrase. The acoustic cues that a speaker produces in such a context have sometimes been referred to as 'maximal' or 'standard'. We adopt the term 'citation', but do not intend this term to signify that these citation form cue patterns are in any way to be preferred (or that other cue sets are somehow undesireable), but rather that they enjoy a certain status as providing substantial evidence for the listener of the features, phonemes and words intended by the speaker. It is also possible that citation cues, particularly landmark cues, play a special role in defining the contrastive phonemic categories of human languages; as Stevens (1972) proposed, their quantal nature may have shaped these categories during the evolution of human language. However, these issues are beyond the scope of the current paper, and will not be dealt with further here.

But how are lexical items represented in the lexicon residing in the human brain, that is, how are lexical items stored in memory? Modeling lexical access requires setting a hypothesis on how words are stored in the mental lexicon. Different theories have been proposed, reflecting different views and beliefs. The Gibsonian tradition, for example, postulates a direct connection from perception to action. This thinking has led to the proposal, in the speech field, that when a human perceives the sound pattern of a spoken utterance, the immediate reaction is to form an representation of how to produce it, and that the elements on which speech perception is based are the intended articulatory gestures of the speaker, represented in the brain as default motor commands (Liberman and Mattingly, 1985; Browman and Goldstein 1985, 1992; Galantucci et al., 2006), the gesture commands being the physical support for abstract phonetic concepts embodied in phonetic features. Since, due to coarticulation, vocal tract gestures temporally overlap in production, the resulting acoustic speech signals are highly context-dependent; perceptual evidence indicates that listeners may recover discrete phonetic gestures from such context-dependent speech signals (Fowler, 2006). While this approach provided significant insights into how listeners make use of articulatory-based perceptual representations, mapping articulatory patterns extracted from acoustic signals, that reflect a large degree of phonetic variation, to lexical contrasts is still a challenging issue. This challenge is particularly acute when modifications observed in running speech are related to the speaker's intended syntax and prosody, i.e. to the structural relationship among the words (Veilleux and Shattuck-Hufnagel, 1998; Moro, 2014), such as phrase final lengthening in English (Turk&Shattuck 2007 journal of phonetics and references therein) and gemination in Italian (Di Benedetto et al., 2021). Moreover, neuropsychological evidence indicates that patients with severe impairments in producing speech gestures are still capable of recognition (Mahon and Caramazza, 2005; Stasenko et al, 2013), which raises questions about the viability of the theory.

Opposed to the gesture approach is the theory of general auditory processing, in which the objects of speech perception are supposed to be purely auditory qualities. This approach suggests that listeners are sensitive to the auditory properties of phonetic events thanks to auditory processes underlying the perception of all sounds, with no requirement for special mechanisms in the perception of speech (Diehl and Kluender, 1989; Kingston et al. 2014). Some aspects of speech sound identification in the face of acoustic variation may be well explained by this approach, although this task only corresponds to the first phase of word identification, and also cannot explain experimental evidence for phonetic decisions dependent upon lexical status (Ganong, 1980).

An alternative approach that unifies articulatory, acoustics, and auditory evidence of human speech perception was proposed by Stevens (2002) in his model of lexical access from distinctive features. The hypothesis in Stevens' model is that words are stored in memory as a hierarchical arrangement of linguistically-inspired distinctive features (Jakobson et al., 1952; Chomsky and Halle, 1968), characterized by correlates in both the acoustic and articulatory domains. Distinctive features assume binary values and describe the contrast between words. For example, the contrast between words "cut" vs. "gut" is described by one feature, reflecting voicing and called [stiff vocal folds], that defines the articulatory action in place when producing the word-initial consonant ([+stiff vocal folds] in "cut" vs. [-stiff vocal folds] in "gut"). On this view, the underlying features govern articulatory gestures and patterns, in order to produce the acoustic goals that signal the abstract symbolic phonemic categories that specify words and distinguish one word from another. Understanding how these mechanisms are connected is at the core of Stevens' model of lexical access from features, since this understanding ultimately provides the critical information that encodes words in the speech signal. The hierarchical organization of features (Clements, 1985) in Stevens' model reflects the postulate that there exist primary stages in the perception, corresponding to the first phase of recognition by a listener, in which abrupt acoustic events - named landmarks - are detected. These landmark cues provide basic information about the sequence of a particular kind of features, i.e. those that do not specify any particular articulator in action during the production, and for this reason are called *articulator-free* features (Halle, 1992). Articulator-free features, also indicated as manner features (reflecting the general concept of "manner of articulation"), classify general classes of speech sounds such as vowels and broad classes of consonants. This set of features includes e.g. [consonantal], [sonorant] and [continuant]. Detecting the abrupt acoustic landmarks that signal the articulator-free features provides an initial estimate of the Consonant-Vowel structure of the utterance. The temporal area around each landmark is then further processed by the listener to extract acoustic cues to the features that are articulator-specific – called *articulator-bound* features –reflecting for example place of articulation (the feature [+anterior] vs. [-anterior] describes the position of the tongue blade constriction against the hard palate, for instance, while the feature [high] describes the height of the tongue body during the production of vowels). These articulator-bound features are also signalled by sets of individual acoustic cues which can vary with context (Huilgol et al., 2019).

The landmark model proposes an innovative vision and a radically different perspective with respect to basic assumptions underlying the design of current automatic speech processing and recognition systems. Current systems are based on short-term speech analysis, and operate by estimating time-domain and frequency-domain acoustic properties on a consecutive and regular frame basis, most often using a frame-synchronous time scale, with vectors of measurements extracted and analyzed at regular and clocked time intervals. The landmark theory offers a profoundly different understanding of what basic human speech processing mechanisms may look like. An illustration of the advantages of this approach is found in a recent article entitled "*Acoustic landmarks contain more information about the phone string than other frames for automatic speech recognition with deep neural network acoustic model*" (He et al. 2018). This paper reports improvement of recognition scores and reduction of computational complexity of a traditional statistically based classifier, when the concept of landmark is introduced, highlighting the idea that certain privileged frames contain more information than others. This result suggests that modeling human speech perception may provide benefits for an engineering perspective. As a matter of fact, current speech recognition systems run in the cloud, because their high computation load would require unaffordable time and battery power to run on current mobile devices (He and Sainath et al., 2019); knowledge-based structural design may help to address this challenge. After all, humans are gifted with this remarkable species-specific ability to acquire any human language and master native language sounds within a year of their birth (Berwick and Chomsky, 2016); reverse engineering this process may prove to have a tremendous impact on both science and technology. Research based on Stevens' approach, that leads to development of a speech recognition system that imitates the mechanism that human listeners employ when identifying words in running speech, may provide a framework to better understand human speech processing in both typical and atypical speakers, shedding light on the mechanisms that are disordered in clinical populations. So far, Stevens' approach has been adopted in the Speech Communication Group of the Massachusetts Institute of Technology (MIT), where research activity is aimed at the development of a lexical access system for English (Stevens, 2002). In this paper, the first steps are taken toward applying this approach in Italian, which will be the first language beyond English to be investigated. Exploring a new language will provide insight into whether Stevens' approach has universal application across languages, with implications for understanding how the human brain recognizes speech.

The paper is organized as follows. Section II provides a synthesis of the principles underlying Stevens' model of Lexical Access, and includes a description of features and acoustic cues to features. Section III contains the description of the Lexical Access system developed for American English. The application of Stevens Lexical Access model to Italian is analyzed in Section IV, which contains a description of the set of phonemes that has been adopted, the proposed features for each phoneme, the Lexical Access corpus for Italian (LaMIT database) and its feature cue labeling. Future research directions and a proposed extension of the model are reported in Section V. Section VI summarizes the earlier sections and concludes the manuscript.

## II. Stevens' Lexical Access model

A theory of lexical representation and lexical access that integrates the knowledge available from the three disciplines of phonology, articulatory phonetics and acoustic phonetics, and takes advantage of their interrelationship, has been developed by Stevens at the Massachusetts Institute of Technology (1992, 1998, 2000, 2002, 2005). To the best of our knowledge no other proposed framework, nor single other approach, provides this wide and comprehensive view of human speech perception. The fundamental hypothesis in Stevens' model is that words are stored in memory as a hierarchical arrangement of linguistically-motivated distinctive features (Jakobson et al., 1952; Chomsky and Halle, 1968), characterized by correlates in both the acoustic and articulatory domains. An excellent reason for using such a feature-based representation is that, when appropriately defined, acoustic properties manifested by a speech signal are in direct relation both to articulatory gestures and to features. The inventory of distinctive features was conceived to be universal (Trubetskoy, 1939; Jakobson et al., 1952), i.e. independent of any particular language, since it was designed to reflect fundamental acoustic-articulatory properties of the human production system. Each language makes use of that subset of features that best describe its language-specific contrasts.

Representing the speech signal of a particular utterance, which is a time-continuous and amplitude-continuous signal, in terms of a sequence of bundles of features involves translation into a discrete representation of the signal in both time and amplitude domains. This process leads to a fully discrete representation. For example, a word is represented as a matrix of "+" and "-" attributes of features. The feature model finds a justification beyond the linguistic classification for which features represent phonological contrasts, since changing the "+" or "-" value of a feature changes the phonological form of a word, which is then no longer associated with the same meaning. The mapping of articulatory dynamics (that evolve in a continuous domain) onto a feature representation (that is by essence discrete) is possible thanks to the quantal nature of speech, as theorized again by Stevens (1972). Stevens showed that a continuous-to-discrete transformation takes place when going from articulatory gestures to acoustic parameters, that is, a change in the configuration of an articulatory structure does not always lead to a change in acoustic parameters. That is, changes within certain ranges of the articulatory parameters may provoke very little change in the acoustics, while for other articulatory ranges the acoustic parameters are quite sensitive. In other words, the articulatory-acoustic relations are quantal. The quantal nature of speech provides, therefore, the theoretical basis for a discrete representation of the information contained in a given word.

But what about time domain, that is, where along the time axis should acoustic parameters be estimated in order to derive acoustic cues to features? Should they somehow be integrated over time? How does human speech perception deal with this interpretation problem? Does it either progressively provide intermediate measurements of cues to then provide at some point an interpretation of cues as evidence for features, or does it rather focus in particular instants of time where the acoustic signal carries particularly relevant information? An answer to these questions can be found in Stevens' Lexical Access model (Stevens, 2002). This model reflects the postulate that there exist primary stages in the perception process, corresponding to the first phase of recognition by a listener, in which abrupt acoustic events - termed landmarks - are detected. In other words, perception does not proceed as a regular sequence of consecutive and systematic measurements over the acoustic signal, but rather is triggered by abrupt or salient acoustic events, i.e. the landmarks. As such, human perception mechanisms can be pictured as a typical hybrid system model (van der Schaft and Schumacher, 2000). In such a model, continuous dynamics describe the continuous operation of the peripheral auditory system, which is always in an awake state, and these continuous dynamics mix with discrete dynamics describing the disruption of regularities in a fully asynchronous manner. These disruptions provoke a change in the continuous system, by pushing it into a different mode of operation in which the system action is to actively perform measurements. Once the relevant information is extracted, the continuous system goes back to its awake but inactive state, waiting for the next trigger to switch again to the active measurement mode. Clearly, while in the awake inactive state, the continuous system does perform some kind of measurement in a continuous manner, since it must be able to detect the triggering event, as does a radio node of a wireless communication system.

According to Stevens' model, these landmark cues provide basic information about the sequence of a particular kind of feature, i.e. those that do not specify any particular articulator in action during production, and for this reason they are called articulator-free features. Articulator-free features, which are roughly equivalent to manner features (reflecting the general concept of "manner of articulation"), classify general classes of speech sounds such as vowels and broad manner classes of consonants. Landmark detection provides, therefore, an initial estimate of the Consonant-Vowel structure of the utterance. The temporal area around each landmark is then further processed by the listener to extract acoustic cues to features that are articulator-specific – called articulator-bound features – reflecting for example place of articulation (for instance the feature [+anterior] vs. [-anterior] describing the position of the tongue blade constriction against the hard palate for instance, or the feature [high] describing the height of the tongue body during the production of vowels). In summary, Stevens' model reflects a hierarchically organized set of features (Clements, 1985).

As mentioned above, articulator-free features classify segments into broad classes, such as vowels and general classes of consonants. When pronouncing a vowel, the vocal tract is open and the air flows freely through the vocal tract. There is no significant narrowing of the vocal tract and this means that there is no significant pressure drop across any narrowing in the system above the glottis (supra-glottal airway). In contrast, consonants are produced by significantly narrowing the airway in the oral cavity. The effect on the acoustic signal of this very basic difference is that vowels are characterized by a much greater intensity of radiated sound than consonants, and by a higher value of the vocal tract first resonance, i.e. higher first formant frequency, due to a greater openness of the vocal tract. The production of consonants provokes discontinuities in the signal, at times that correspond to the consonant closure and the consonant release. The closure can be partial, as in continuant consonants, or complete, as in the production of obstruents, at least in citation form. Consonants typically differ from vowels by the presence of these discontinuities and also by the lower amplitude of the spectrum in the low and mid frequency regions.

As a matter of fact, a particular class of sounds – glides –, are described as neither vowels nor consonants, since they are characterized by reduced amplitude in the low and mid frequency spectrum regions compared to vowels, but this effect is obtained without generating a constriction. Glides are sometimes described as semi-vowels or semi-consonants, to express the above "mixed" behavior. In summary, the detection of acoustic cues to a sequence of articulator-free features provides an initial estimate of the structure of an utterance as a sequence of vowels, true consonants, and glides[2]. The corresponding articulator-free features are [vowel], [consonant], and [glide]. In the case of consonantal segments, additional articulator-free features characterize the manner of articulation during the production, corresponding to: 1) the feature [continuant] (for example fricatives are [+continuant] since the closure is only partial, while stops are [-continuant] because closure is complete); 2) the feature [sonorant] (this feature is only specified for [+consonant] segments and is positive if there is no increase in pressure behind the consonant closure, as in nasals); 3) the feature [strident] (only defined for [+continuant] segments, to distinguish segments with a high amplitude in the higher spectral frequencies, that is much greater than that of adjacent vowels, i.e. the [+strident] segments, from segments that do not have this property).

---

[2] Of course not all landmarks will be realized in all utterances but the hypothesis is that enough landmarks will be realized to enable accurate perception in context.

Regarding articulator-bound features, there are seven articulators that can be manipulated to produce phonetic distinctions of a language. These articulators activate during the production of the vowel-consonant-glide structure induced by the articulator-free features and are manipulated to produce contrasts corresponding to distinctive features of a language. Figure 2 shows the seven articulators and corresponding features as proposed in Keyser and Stevens (1994). The articulators are grouped into four sets: articulators in the oral cavity (lips, tongue blade and tongue body), articulators in the pharyngeal and laryngeal regions that control the shape of the [back half of the?] vocal tract (larynx, pharynx, and glottis), articulators that couple the nasal cavity (raising and lowering of the soft palate), and articulators that adjust the stiffness or slackness of the vocal folds, without affecting the shape of the vocal tract (vocal folds). A language uses a set of those articulator-free and articulator-bound features to express contrasts between words in its lexicon. Not all features defined in Fig. 2 are distinctive in fact in a given language, and the proposed set may not be exhaustive for all languages; It may need to be expanded.

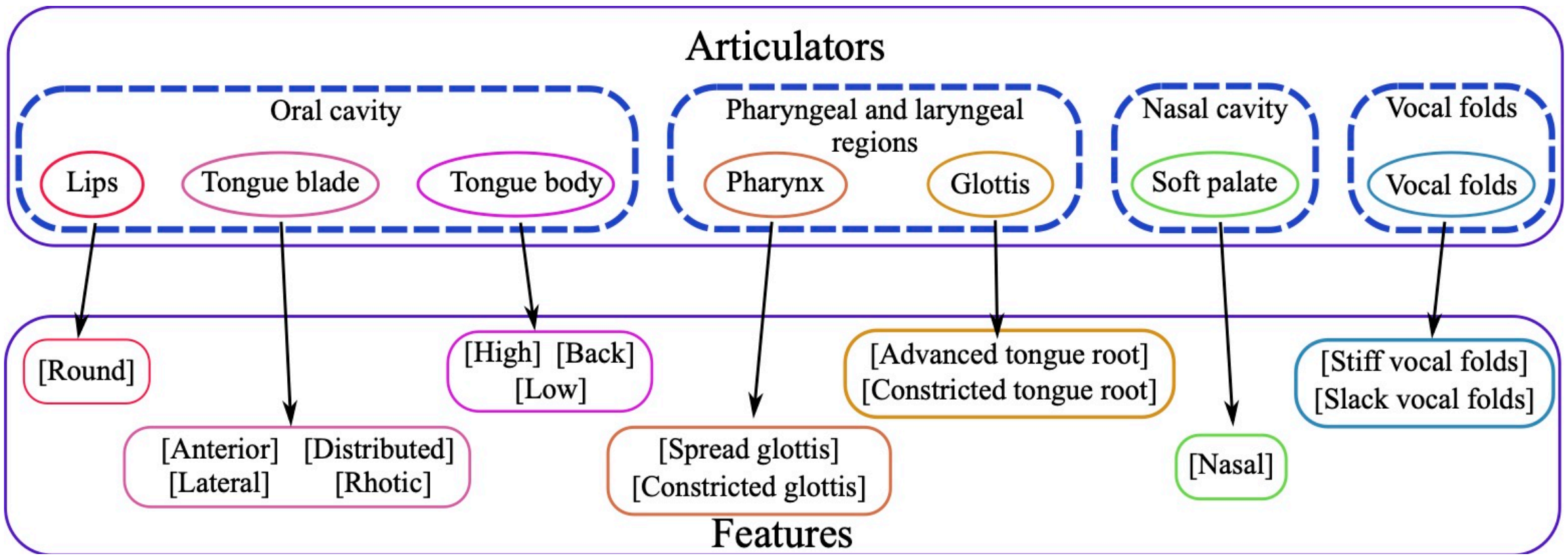

Figure 2 - The seven articulators and corresponding features proposed by Keyser and Stevens (1994).

In summary, Stevens' Lexical Access model proposes an innovative perspective with respect to basic assumptions underlying the design of the human speech production and recognition systems. These systems are based on short-term speech analysis and operate by estimating time-domain and frequency-domain acoustic properties on a consecutive and regular frame basis, most often using a frame-synchronous time scale, with vectors of measurements extracted and analyzed at regular and clocked time intervals. The landmark theory offers a distinct perspective to what basic human speech processing mechanisms may look like. Most interestingly, very recent findings in the field of neurophysiology, obtained thanks to the tremendous advances in functional hemodynamic neuroimaging and electrophysiological methods, report evidence for a hierarchical feature representation in the Superior Temporal Gyrus (STG). In particular, Oganian and Chang (2019) report a specific area of the cortex of the human brain, in the medial STG, that responds strongly to a specific characteristic of speech amplitude envelope: the acoustic onset edge, such as between a consonant and a vowel (as in "see"), or between a fricative consonant and a stop consonant (as in "spa"). This sensitivity to abrupt changes in the speech envelope provides strong support for Stevens' landmark-based theory of speech perception, since landmarks are the abrupt acoustic changes that signal certain basic distinctive features of the speaker's intended sequence of phonological categories, the articulator-free manner features. Moreover, evidence for a hierarchical feature representation in the STG has been documented based on direct cortical recordings (Chang et al. 2010). These findings support the hypothesis of an acoustic-phonetic representation of heard speech in the human brain, that follows a systematic organization consistent with models of human perception in which distinctions are primarily driven by manner contrasts. Thus these recent neurophysiological findings provide additional evidence and support for the landmark-based approach that is embodied in Stevens' model (Mesgarani et al., 2014).

## III. The American-English Lexical Access system

The Lexical Access system developed at MIT - for spoken English - over a span of more than 20 years, consists of a complex association of modules dedicated to different system functions, such as detection of landmarks, interpretation of specific features, and ultimately word recognition (Huilgol et al., 2019). The detection modules are based on the definition of feature sets for vowels, consonants and glides, summarized here in Table I for vowels and glides, and Table II for consonants (Choi, 2012).

Table I – Standard feature sets for vowels and glides

| IPA | i ɪ<br>iy ih | e ɛ<br>ey eh | æ<br>ae | a ɔ<br>aa ao | o ʌ<br>ow ah | u ʊ<br>uw uh | ɚ ə<br>rr ex | aʊ<br>au | aɪ<br>ai | ɔɪ<br>oi | h w j ɹ l<br>h w y r l |
|---|---|---|---|---|---|---|---|---|---|---|---|
| Vowel | + + | + + | + | + + | + + | + + | + + | + | + | + | |
| Glide | | | | | | | | + | + | + | + + + + + |
| Cons | | | | | | | | | | | |
| Son | | | | | | | | | | | |
| Cont | | | | | | | | | | | |
| Strid | | | | | | | | | | | |
| Stiff | | | | | | | | | | | |
| Slack | | | | | | | | | | | |
| Spread | | | | | | | | | | | |
| Const | | | | | | | | | | | |
| Nasal | | | | | | | | | | | |
| Blade | | | | | | | | | | | |
| Body | | | | | | | | | | | |
| Lips | | | | | | | | | | | |
| Atr | + - | + - | + | + - | + - | + - | - - | - - | - - | - - | + + - |
| Ctr | | | | | | | | | | | |
| High | + + | - - | - | - - | - - | + + | - - | - + | - + | - + | + + - - |
| Low | - - | - - | + | + + | - - | - - | - - | + - | + - | + - | - - - - |

| | | | | | | | | | | |
|---|---|---|---|---|---|---|---|---|---|---|
| Back | - - | - - | - | + + | + + | + + | + + | + + | + - | + - | + - + + |
| Round | | | | + | + | + + | | + | | + | + - - |
| Ant<br>Dist<br>Lat<br>Rhot | | | | | | | -<br>-<br>-<br>+ | | -<br>+ | -<br>+ | - - +<br>+ - -<br>- +<br>+ - |

| Table II – Standard feature sets for consonants | | | | | | |
|---|---|---|---|---|---|---|
| IPA Label | m n ŋ<br>m n ng | v ð z ʒ<br>v dh z zh | f θ s ʃ<br>f th s sh | b d g<br>b d g | p t k<br>p t k | dʒ tʃ<br>dj ch |
| Vowel<br>Glide<br>Cons<br>Son<br>Cont<br>Strid | <br><br>+ + +<br>+ + +<br>- - - | <br><br>+ + + +<br>- - - -<br>+ + + +<br>- + + | <br><br>+ + + +<br>- - - -<br>+ + + +<br>- + + | <br><br>+ + +<br>- - -<br>- - - | <br><br>+ + +<br>- - -<br>- - - | <br><br>+ +<br>- -<br>± ±<br>+ + |
| Stiff<br>Slack | | <br>+ + + + | + + + + | <br>+ + + | + + + | +<br>+ |
| Spread<br>Const | | | | | | |
| Nasal | + + + | | | | | |
| Blade<br>Body<br>Lips | +<br>+<br>+ | <br>+ + +<br>+ | <br>+ + +<br>+ | +<br>+<br>+ | +<br>+<br>+ | <br>+ + |
| Atr<br>Ctr | | | | | | |
| High<br>Low | +<br>- | | | +<br>- | +<br>- | |
| Back | + | | | + | + | |
| Round | - | - | - | - | - | |
| Ant<br>Dist<br>Lat<br>Rhot | +<br>- | + + -<br>+ - + | + + -<br>+ - + | +<br>- | +<br>- | - -<br>+ + |

A first step in analyzing a spoken utterance for lexical access is to detect all three types of landmarks: vowel landmarks, glide landmarks, and consonant landmarks. A second step is to extract the articulator-bound features based on acoustic cues that are evaluated around the landmarks. As noted above, vowel landmarks correspond to places in the signal where the amplitude of low-frequency regions in the spectrum is maximum, or where the first formant (F1) reaches it maximum, indicating a maximum opening above the glottis. Consonant landmarks correspond to places where abrupt discontinuities in the signal are observed, signaling constrictions and releases of the oral vocal tract. Finally, glide landmarks correspond to minima in the low-frequency amplitude, adjacent to vowel landmarks, and signaling an extreme narrowing in the airway that is not constricted enough to cause an acoustic discontinuity.

Landmark detection was extensively analyzed, and in particular a consonant detection module was completed by Liu (1995, 1996), to identify the time at which constrictions are created and released. The detector also characterizes the landmark as [+sonorant], [+continuant] (as in fricatives), or [-continuant] (as in stops). A vowel detection module was proposed by Howitt (2000), operating on peaks in the low-frequency amplitude. Glide detection was addressed by Sun (1996), for the particular case of /w/ and /j/. Choi (1999) classified consonant voicing, and Lee and Choi (2008a,b) and Lee et al. (2011, 2012) classified place of articulation. More recently, acoustic landmarks have been used in automatic speech recognition systems (see Hasegawa-Johnson et al., 2005; Juneja, 2004, Kong et al., 2016).

According to the Stevens model, acoustic cues to articulator-bound features must in some way relate to articulatory gestures that are produced by the speaker to signal a contrast defined by a distinctive feature. Most important cues must reflect properties that are speaker-independent. Stevens (1998) reports an inventory of the acoustic parameters that provide evidence for relevant configurations and movements of the articulatory structures, and in particular:

1. Parameters that are related to the position of tongue body and to lip rounding
2. Parameters that signal within vocalic regions the presence of a velopharyngeal opening
3. Parameters that signal in the spectrum a rapid movement of the articulators, in particular tongue blade and lips
4. Parameters describing vocal-fold stiffness and in particular frequency of vibration at the glottis
5. Parameters that describe the state of the glottis in a vocalic region
6. Parameters describing the place of articulation during frication noise
7. Parameters that describe the state of the glottis when the supraglottal pressure is increased
8. Parameters that signal a velopharyngeal opening within constricted regions and during the nasal murmur
9. Parameters providing information on subglottal pressure
10. Parameters providing information about the distance in time between two landmarks

The first five categories correspond to temporal regions during an utterance where the vocal tract is not constricted, as in vocalic regions, and the sound source is at the glottis. Parameters in categories 6 to 8 refer to information about the supraglottal and pharyngeal states in regions where there is a consonantal constriction. Parameters of category 9 signal the beginning and the end of a sentence, while parameters of category 10 are related to temporal characteristics of a sentence.

In the vicinity of vowel and glide landmarks, the acoustic parameters that should be considered are related to categories 1, 2, 4, and 5. In the vicinity of consonant landmarks, parameters should be measured in the adjacent vocalic regions (categories 1, 2, 3, 4 and 5) and in the constricted regions (categories 6, 7, 8). In the MIT system, both landmarks and parameters are referred to as acoustic cues to distinctive features.

## IV. Application of Stevens' Lexical Access model to the Italian language

The American-English system is at the core of the initial work on this approach to lexical access, which forms the basis for the development of the Lexical Access system for the Italian language.

The LaMIT database (acoustic materials and labeling information) can be accessed freely at the following address: http://acts.ing.uniroma1.it/project_lamit_database.php under an open-source Creative Commons license.

### IV.1 The adopted Italian phoneme set

We adopt the phoneme set proposed by Muliacic (1972) to represent the phonemes of the Italian language. Following Muljacic, this study considered all vowels monophthongs. The adopted phonemic classification was thus the seven vowels: /a/, /i/, /u/, /e/, /ɛ/, /o/, /ɔ/; the twenty-one consonants: /p/, /b/, /f/, /v/, /t/, /d/, /ts/, /dz/, /s/, /z/, /k/, /g/, / dʒ/, /tʃ/, /ʃ/, /m/, /n/, /ɲ/, /l/, /ʎ/, /r/; and the two glides: /j/ and /w/. Vowel-glide and glide-vowel segments were not treated as diphthongs and were instead assumed to contain two distinct phonemes. This set was enriched with consonants in their geminated form. As a matter of fact, most Italian consonants can be geminated, with the exception of a few such as /z/, although different views are expressed also regarding a particular subset of five consonants /ts, dz, ʃ, ɲ, ʎ /, about which contrasting views are found in the literature. Note that gemination is word contrastive in Italian, i.e. distinguishes between word pairs like "pala" (shovel) vs. "palla" (ball) – a property shared by few languages; an exhaustive analysis of lexical gemination carried out on a corpus of VCV vs. VCCV words can be found in (Esposito and Di Benedetto, 1999), (Di Benedetto and De Nardis 2021a), (Di Benedetto and De Nardis 2021b); a description of the corpus itself and information on how to obtain it can be found in (Di Benedetto and De Nardis 2021c). In Italian, gemination may also be elicited by certain syntactic contexts, and in this case, it involves junction and adjustment between consonants occurring at word boundaries, giving rise to syntactic gemination (*Raddoppiamento Sintattico*). Recent studies (Di Benedetto et al., 2021) further investigated two opposite views on this issue, that is whether consonant gemination is the process by which a consonant is produced i) as two consecutive occurrences of a same phoneme, or, ii) as a stronger, longer (or sometimes indicated as more intense) version of the consonant. The study was carried out on the acoustic materials provided by the LaMIT database (fully described in Section IV.3) and was also motivated by questions raised during the development of the present framework; It proceeded in parallel to the development of the model presented in this paper. Evidence was found that in **running speech**, a geminated consonant is the result of the lexical doubling of single consonantal phonemes, rather than of intensifying the consonant. This was observed for both lexical and syntactic geminates. In the case of stop consonants, for instance, a substantial number of instances of double closures and bursts were observed, providing evidence for the presence of two consecutive consonants. Such double bursts were found in about 12% of instances of both lexical and syntactic geminates (Di Benedetto et al., 2021). On this basis, geminate consonants were considered in this study as two consonants, both for lexical and syntactic germinated consonants.

Based on the above observations, the finalized set of Italian phonemes is as shown in Table III, that also shows the "computer-friendly" ARPAbet notation of each Italian phoneme.

Table III – List of Italian phonemes and of their ARPAbet notation

| IPA Phoneme | ARPAbet Phoneme | IPA Phoneme | ARPAbet Phoneme |
|---|---|---|---|
| /a/ | AA | /k/ | K |
| /e/ | EY | /kk/ | KK |
| /ɛ/ | EH | /g/ | G |
| /o/ | OW | /gg/ | GG |
| /ɔ/ | AO | /t/ | T |
| /i/ | IY | /tt/ | TT |
| /u/ | UW | /d/ | D |
| /l/ | L | /dd/ | DD |
| /ll/ | LL | /f/ | F |
| /ʎ/ | LH | /ff/ | FF |
| /ʎʎ / | LHLH | /v/ | V |
| /r/ | R | /vv/ | VV |
| /rr/ | RR | /s/ | S |
| /j/ | Y | /ss/ | SS |
| /w/ | W | /z/ | Z |
| /n/ | N | /ʃ/ | SH |
| /nn/ | NN | /ʃʃ/ | SHSH |
| /m/ | M | /tʃ/ | CH |
| /mm/ | MM | /tʃtʃ/ | CHCH |
| /ɲ/ | GN | /dʒ/ | JH |
| /ɲɲ/ | GNGN | /dʒdʒ/ | JHJH |
| /p/ | P | /ts/ | TS |
| /pp/ | PP | /tsts/ | TSTS |
| /b/ | B | /dz/ | DZ |
| /bb/ | BB | /dzdz/ | DZDZ |

**IV.2 Feature sets of Italian vowels and consonants**

The set of features defined for English served as an initial ground for defining features for Italian, whenever possible, and in particular for those segments that share, in Italian and English, common properties in terms of articulatory gestures involved during the production. The next task was to devise an original feature inventory for phonemes that are present in the Italian lexicon but not in English, i.e. /ʎ/, /ɲ/, /ts/, and /dz/. The phoneme /ɲ/, for example, bears almost all the features of /n/, but is articulated with the tongue body rather than the blade. Thus, it receives the features [+body] [-blade]. The phoneme /ʎ/ resembles /l/ but is more dorsal; accordingly, it is classified as [-ant] [+dist] and [-blade] [+body]. Like all dorsal phonemes, it is also classified by height: [+high] [-low] [-back]. Finally, the phoneme /dz/ resembles /ts/ but is voiced; it is assigned the same features as /ts/ but is [+slack] (while [ts] is [+stiff]). Both /ts/ and /dz/ receive the same features as the English /dʒ/ and /tʃ/ except for the feature [anterior]: /ts/ and /dz/ are [+ant] while /dʒ/ and /tʃ/ are [-ant].

The proposed feature attributes for these phonemes are in line with Hayes (2009). Hayes's registry of features aims chiefly to categorize the phonemes of every language by subsets of "natural classes". Hayes's "manner" features are analogous to Stevens' "articulator free" features, while "place" features map on to "articulator bound" features.

As a side note, the Italian rhotic (/r/) is a coronal trill but may manifest as an allophonic flap or tap. It was not necessary to include /ɾ/ in this inventory because there are no lexical contexts in which an /r/ to /ɾ/ lenition would occur paradigmatically; it is an idiosyncrasy of the speaker that may, however, be connected to vowel lengthening. Though Stevens does not include the features [tap] or [trill], /r/ can be uniquely defined in relation to other coronal consonants using extant features.

In summary, Table IV shows the complete set of distinctive features of Italian vowels, glides, and consonants.

Table IV – Distinctive features set for the Italian vowels, glides, and consonants

| | a | e | i | o | u | ɛ | ɔ | j | w | p | b | f | v | t | d | ts | dz | s | r | z | k | g | dʒ | tʃ | ʃ | m | n | ɲ | l | ʎ |
|---|---|---|---|---|---|---|---|---|---|---|---|---|---|---|---|---|---|---|---|---|---|---|---|---|---|---|---|---|---|---|
| Vowel | + | + | + | + | + | + | + | | | | | | | | | | | | | | | | | | | | | | | |
| Glide | | | | | | | | + | + | | | | | | | | | | | | | | | | | | | | | |
| Cons | | | | | | | | | | + | + | + | + | + | + | + | + | + | + | + | + | + | + | + | + | + | + | + | + | + |
| Cont | | | | | | | | | | - | - | + | + | - | - | ± | ± | + | + | + | - | - | ± | ± | + | + | + | + | + | + |
| Son | | | | | | | | | | - | - | - | - | - | - | - | - | - | + | - | - | - | - | - | - | + | + | + | + | + |
| Strid | | | | | | | | | | | | | | | | + | + | + | - | + | - | - | + | + | + | - | - | - | - | - |
| Lips | | | | | | | | | | + | + | + | + | | | | | | | | | | | | | + | | | | |
| Blade | | | | | | | | | | | | | | + | + | + | + | + | + | | | | | | + | | + | | + | |
| Body | | | | | | | | + | | | | | | | | | | | | | | | | | | | | + | | + |
| Round | | | | | | | | | + | - | - | | | | | | | | | | | | | | | - | | | | |
| Ant | | | | | | | | | | | | | | + | + | + | + | + | + | + | | | - | - | - | | - | - | + | - |
| Dist | | | | | | | | | | | | | | | | | | | | - | | | + | + | + | | | | - | + |
| Lat | | | | | | | | | | | | | | | | | | | | | | | | | | | | | + | + |
| High | - | - | + | - | + | - | - | + | + | | | | | | | | | | | | | | | | | | | | | + |
| Low | + | - | - | - | - | - | - | - | - | | | | | | | | | | | | | | | | | | | | | - |
| Back | - | - | - | + | + | - | + | - | + | | | | | | | | | | | | | | | | | | | | | - |
| ATR | + | + | + | + | + | | | + | + | | | | | | | | | | | | | | | | | | | | | |
| CTR | | | | | | + | + | | | | | | | | | | | | | | | | | | | | | | | |
| Spread glottis | | | | | | | | | | | | | | | | | | | | | | | | | | | | | | |
| Constr glottis | | | | | | | | | | | | | | | | | | | | | | | | | | | | | | |
| Nasal | | | | | | | | | | | | | | | | | | | | | | | | | | + | + | + | | |
| Stiff | | | | | | | | | | + | | + | | + | | + | | + | | | + | | | + | + | | | | | |
| Slack | | | | | | | | | | | + | | + | | + | | + | | + | + | | + | + | | | | | | | |

### IV.3 The proposed Lexical Access Italian corpus: the LaMIT database

A corpus of speech data for Italian was recorded, to form the LaMIT database. The LaMIT database will provide the speech community at large with a database of sentences uttered by several speakers, to serve as a reference database comparable to the TIMIT database (Garofolo et al., 1993), a widely used fully transcribed database for American-English that was created for training speech recognition software and consists of about 10 sentences uttered by over 600 speakers. A corpus that reflects the richness in phonemic representation in the language is essential to implementing and testing the Lexical Access model. Thus, similar in size to the LAFF sentences database created for the American-English language (Speech communication group (MIT), 2005), the LaMIT database includes 100 sentences, uttered by 2 male speakers (LDN and JV) aged 40 and 20, respectively, and 2 female speakers (MGDB and SB) aged 60 and 24, respectively, with 2 repetitions of each sentence. A recent study (Arango et al., 2021) that was carried out in order to finalize the LaMIT corpus, provided updated values of the phonemic frequencies; of particular interest is the fact that, unlike previous analyses of similar kind (for example

Busa et al., 1962) this study also reports frequency of occurrence for geminated consonants. Table V contains the phonemic frequency of the 30 Italian phoneme and of the 20 geminated Italian consonants. Based on the above data on frequency of occurrence, the LaMIT database was tailored so to reflect the typical frequency of occurrence of the different phonemes and of their geminated forms. Table VI reports the 100 sentences of the LaMIT database.

Table V - Phonemic frequencies in LaMIT vs. Arango et al. (2021)- Left: frequency of each of the 30 IPA phonemes in singleton form; right: frequency of each of the 20 geminated IPA phonemes.

| **Singleton** | | | | | | **Geminated** | | | | | |
|---|---|---|---|---|---|---|---|---|---|---|---|
| **Phoneme** | **Frequency %** | | **Phoneme** | **Frequency %** | | **Phoneme** | **Frequency %** | | **Phoneme** | **Frequency %** | |
| | Arango et al. (2021) | LaMIT | | Arango et al. (2021) | LaMIT | | Arango et al. (2021) | LaMIT | | Arango et al. (2021) | LaMIT |
| /a/ | 11.92 | 12.99 | /v/ | 1.83 | 1.89 | /ll/ | 1.09 | 0.96 | /mm/ | 0.08 | 0.1 |
| /e/ | 10.07 | 9.68 | /j/ | 1.80 | 1.87 | /tt/ | 0.91 | 0.82 | /dzdz/ | 0.03 | 0.1 |
| /i/ | 9.17 | 9.03 | /w/ | 1.16 | 0.86 | /ss/ | 0.53 | 0.43 | /vv/ | 0.03 | 0.02 |
| /o/ | 8.71 | 8.2 | /ɔ/ | 1.08 | 1.13 | /tsts/ | 0.42 | 0.31 | /dd/ | 0.01 | 0.02 |
| /n/ | 7.32 | 7.17 | /tʃ/ | 0.86 | 0.72 | /dʒdʒ/ | 0.24 | 0.26 | /gg/ | <0.01 | <0.01 |
| /r/ | 6.42 | 6.64 | /g/ | 0.76 | 0.72 | /bb/ | 0.23 | 0.12 | | | |
| /t/ | 4.99 | 5.10 | /f/ | 0.74 | 0.89 | /kk/ | 0.21 | 0.31 | | | |
| /l/ | 4.29 | 4.96 | /b/ | 0.71 | 1.03 | /ɲɲ/ | 0.19 | 0.12 | | | |
| /s/ | 4.21 | 4.24 | /dʒ/ | 0.38 | 0.74 | /nn/ | 0.18 | 0.24 | | | |
| /d/ | 4.21 | 3.71 | /ts/ | 0.34 | 0.29 | /rr/ | 0.17 | 0.14 | | | |
| /k/ | 3.83 | 2.95 | /ʎ/ | 0.06 | 0.02 | /pp/ | 0.15 | 0.17 | | | |
| /p/ | 2.88 | 2.80 | /ʃ/ | 0.06 | 0.12 | /ʎʎ / | 0.14 | 0.19 | | | |
| /u/ | 2.52 | 2.66 | /z/ | 0.04 | 0.14 | /ʃʃ/ | 0.12 | 0.12 | | | |
| /m/ | 2.47 | 2.49 | /dz/ | 0.03 | 0.02 | /ff/ | 0.08 | 0.12 | | | |
| /ɛ/ | 2.24 | 2.28 | /ɲ/ | <0.01 | 0.02 | /tʃtʃ/ | 0.08 | 0.07 | | | |

Table VI – Sentences of the LaMIT database

1. Il gatto della vicina è bianco peloso e pazzo
2. Il giardino di mio cugino è pieno di gladioli e di gnomi
3. L'università italiana è un'istituzione pubblica dello stato
4. Passeggerei volentieri a piedi nudi nella città vecchia
5. Pietro non scappa fugge a gambe levate con il cuore in fiamme
6. Cosa ne penseresti di alzarti presto e salutare il sole
7. Quando Maria è in vacanza compra volentieri la settimana enigmistica
8. Lo schermo del tuo cellulare è graffiato e opacizzato
9. All'imbrunire la cattedrale svetta nel cielo basso e uggioso
10. Alcuni studenti dell'anno accademico corrente potranno laurearsi a luglio
11. Due sorelle si aiutano se vanno d'amore e d'accordo
12. La struttura precaria resse malgrado il forte vento
13. Mandare cartoline da città remote non è più di moda
14. Discendi il Monte Bianco con gli sci e vivi un'esperienza unica e indimenticabile
15. Prima o poi dovrai pur deciderti a leggere le opere di Niccolò Machiavelli
16. Non potendo fare a meno del cioccolato pensò bene di privarsi della panna montata
17. Che avventura meravigliosa quella di guardare gattonare un bebè
18. "E pur si muove" disse il famoso scienziato rivolgendosi agli inquisitori
19. Oggi piove a dirotto governo ladro
20. Riporre tanti sogni nel cassetto rinforza la fantasia del poeta
21. Vent'anni di allenamento non furono sufficienti a chiudere la pinza
22. Senti un po' di musica e vedi che ti passa la nostalgia dell'inverno
23. I clienti della Banca devono attenersi alle regole stabilite dal contratto
24. Apponi la firma in calce perché è necessario per rendere valida la transazione
48. Impariamo a meditare giornalmente
49. Si perde così tanto tempo a discutere del niente
50. Stasera andremo al cinema a vedere un film francese
51. Sono belli i programmi decisi all'ultimo momento
52. Con Cristiana pratico yoga ogni mercoledì
53. Pensieri e parole cantava la diva con voce suadente
54. Aprile si esaurisce mentre arriva carico di promesse il mese di maggio
55. Abbiamo trasmesso il giornale radio del mattino
56. Il tempo previsto sull'Italia per questa sera non prevede temperature in aumento
57. Pensavo che tu volessi fare solo uno spuntino
58. Assicurati che non si dimentichino di scrivere alla zia
59. Per salvarci dobbiamo restare uniti
60. Il mondo è nelle nostre mani
61. Comportati educatamente a tavola
62. Pare che sia rimasto solo per un colpo di testa
63. La piccola peste vuole il ciuccio per calmarsi
64. Non mordere la spalla della nonna
65. La carta non si mangia se non sei una capra
66. Il pavone becca le foglie sul viale dello zoo
67. All'improvviso si udì l'urlo del barbagianni
68. Basta con i fanatismi esagerati
69. Non smettere di fantasticare ad occhi aperti
70. Col vento in poppa attraversarono il Mediterraneo in un soffio
71. Voltati e renditi conto di quanta strada hai percorso
72. Una tazza di te verde al giorno rinfresca la mente
73. Scriverò questa lettera con la penna a sfera
74. Che ne pensi di una fetta di torta
75. Il cestino per la carta sta sotto la scrivania

25. I ragazzi della scuola religiosa fisseranno un appuntamento con il sindaco ateo
26. Se prendi in prestito un libro alla biblioteca godi del vantaggio di non dover acquistarlo
27. La rappresentazione digitale delle immagini ha rivoluzionato la fotografia
28. Addio all'imperatore giapponese abdicherà oggi in favore di suo figlio
29. Uno sciame di api investì il bambino biondo costringendolo a buttarsi giù dall'albero
30. La ferrovia si snoda lungo il fiume seguendo un tracciato tortuoso
31. Dopo avere letto molti libri Luca si rimise a studiare ancora per un po'
32. Se arrivi all'alba a Capri butta l'ancora e prosegui a nuoto
33. Giorgio ha deciso di prendere i voti ma prima ha dovuto battezzarsi
34. Che ne farai dei quaderni di storia
35. Chiedi pure a tuo padre cosa ne pensa dell'anguria
36. Mamma e papà ti vogliono bene
37. Non poggiare il bicchiere colmo d'acqua sul pianoforte
38. Con la bicicletta elettrica le salite sono una passeggiata
39. Saluta la signora e fai l'inchino
40. La teoria dei numeri è una branca della matematica
41. Mio nipote ama trovare soluzioni a problemi complessi
42. Aguzza l'ingegno e progetta una radio intelligente
43. Il giornalaio vende e invia riviste e oggetti turistici
44. Il cane corse forsennatamente verso il padrone calpestando le aiuole
45. Il dolore sorgeva mentre la luna non era ancora tramontata
46. Puoi accendere la radio a caso e sintonizzarti su qualsiasi frequenza
47. Poi ci sono i rimedi naturali che sono più efficaci di tanti prodotti presenti in farmacia
76. Una vacanza in agriturismo in Toscana ha un costo ridotto
77. Sul pavimento del salone giace un tappeto persiano
78. L'albero di cedro è simbolo del Libano
79. Un biglietto di auguri accompagna il regalo
80. Torneresti a casa a piedi
81. Il grano saraceno non contiene glutine
82. Il pane lievita quando la luna è piena
83. Stendi il bucato al sole e risparmi energia
84. Sotto la piazza giace un tesoro
85. La balena blu nuota in solitario
86. Servono nuovi dirigenti per rilanciare le aziende
87. Creare lavoro è un dovere costituzionale
88. Ma questa è un'altra storia su cui si indagherà
89. Si al regolamento che impone limiti alla stupidità
90. La ballerina indossa un costume rosa fragola
91. L'autore si muove con scioltezza nella palude delle parole
92. La fascetta giusta dovrebbe essere alienazione
93. Resti in collegamento che risponderà il primo operatore libero
94. Vediamo se la risposta è quella giusta
95. L'avocado cresce nei paesi tropicali
96. Pesce fritto e insalata mista grazie
97. Favorisce un caffè dopo cena col digestivo
98. La folla era impazzita alla vista dell'assassino
99. Lei col maglione rosso si stia zitto
100. Mangerebbe volentieri un filetto di baccalà con le olive

Following the hypotheses laid out above (see for example syntactic gemination) the set of sentences was phonemically transcribed as reported in Table VII. Note that syntactic doubling was annotated as well. Intervocalic /s/ was phonemically labeled as voiceless as suggested by Muljacic (1972)[3].

Table VII – Phonemic transcription of the LaMIT database

| | |
|---|---|
| 1. | 'il 'gatto 'della viˈʧina ˈɛ bˈbjanko peˈloso 'e p'patstso |
| 2. | 'il ʤarˈdino ˈdi ˈmio kuˈʤino ˈɛ pˈpjɛno ˈdi glaˈdioli 'e ɲˈɲɔmi |
| 3. | 'luniversi'ta itaˈljana ˈɛ 'unistitutsˈtsjone ˈpubblika ˈdello ˈstato |
| 4. | passeʤʤɛ'rɛj volenˈtjɛri 'a pˈpjɛdi ˈnudi ˈnella ʧitˈta vˈvɛkkja |
| 5. | ˈpjɛtro ˈnon 'skappa 'fuddʒʤe 'a gˈgambe leˈvate ˈkon 'il ˈkwɔre ˈin 'fjamme |
| 6. | ˈkɔsa ˈne pense'resti ˈdi alˈtsarti ˈprɛsto 'e ssaluˈtare 'il ˈsole |
| 7. | ˈkwando maˈria ˈɛ 'in vaˈkantsa 'kompra volenˈtjɛri ˈla settiˈmana enigˈmistika |
| 8. | Lo ˈskermo ˈdel ˈtuo ʧelluˈlare ˈɛ ggrafˈfjato 'e opaʧidzˈdzato |
| 9. | 'allimbru'nire ˈla katteˈdrale 'zvetta ˈnel ˈʧɛlo 'basso 'e uʤˈʤoso |
| 10. | alˈkuni stuˈdɛnti 'dell'anno akkaˈdɛmiko korˈrɛnte po'tranno lawreˈarsi 'a lˈluʎʎo |
| 11. | ˈdue soˈrɛlle 'si a'jutano ˈse v'vanno da'more 'e ddak'kɔrdo |
| 12. | 'la strutˈtura preˈkarja ˈrɛsse malˈgrado 'il ˈfɔrte ˈvɛnto |
| 13. | manˈdare kartoˈline ˈda ʧitˈta reˈmɔte ˈnon ˈɛ pˈpju 'ddi ˈmɔda |
| 14. | diʃˈʃendi 'il ˈmonte ˈbjanko ˈkon ˈʎi ˈʃi 'e ˈvivi 'unespeˈrjɛntsa ˈunika 'e indimentiˈkabile |
| 15. | ˈprima ˈo pˈpɔj dovr'aj ˈpur deˈʧiderti 'a ˈlɛddʒere ˈle ˈɔpere 'di nikko'lo makja'velli |
| 16. | ˈnon poˈtɛndo ˈfare 'a mˈmeno ˈdel ʧokkoˈlato penˈsɔ bˈbɛne 'di priˈvarsi 'della ˈpanna monˈtata |
| 17. | ˈke avvenˈtura meraviʎˈʎosa ˈkwella 'di gwarˈdare gattoˈnare ˈun beˈbɛ |
| 18. | 'e pˈpur 'si ˈmwɔve 'disse 'il faˈmoso ʃenˈtsjato rivol'ʤɛndosi 'aʎʎi inkwisiˈtori |
| 19. | ˈɔʤʤi ˈpjɔve 'a ddiˈrotto goˈvɛrno ˈladro |
| 20. | riˈporre ˈtanti ˈsoɲɲi ˈnel kasˈsetto rinˈfɔrtsa 'la fantaˈsia ˈdel poˈɛta |
| 21. | 'vent'anni 'di allenaˈmento ˈnon 'furono suffiˈʧɛnti 'a kˈkjudere 'la 'pintsa |
| 22. | 'senti 'un ˈpɔ d'di ˈmusika 'e v'vedi ˈke tˈti ˈpassa 'la nostalˈʤia 'dellin'vɛrno |
| 23. | 'i kliˈɛnti 'della ˈbanka 'dɛvono atteˈnersi 'alle ˈrɛgole stabiˈlite ˈdal konˈtratto |
| 24. | apˈponi 'la ˈfirma 'in ˈkalʧe perˈke ˈɛ nneʧesˈsarjo ˈper ˈrɛndere ˈvalida 'la transatˈtsjone |
| 25. | 'i raˈgatstsi 'della ˈskwɔla reliˈʤosa fisse'ranno 'un appuntaˈmento ˈkon 'il ˈsindako ˈateo |
| 26. | ˈse pˈprɛndi 'in ˈprɛstito 'un ˈlibro 'alla bibljoˈtɛka ˈgɔdi ˈdel vanˈtaʤʤo 'di ˈnon doˈver akkwiˈstarlo |
| 27. | 'la rappresentatˈtsjone diʤiˈtale 'delle imˈmaʤini 'ʔa rivolutstsjoˈnato 'la fotogra'fia |
| 28. | ad'dio allimpera'tore ʤappoˈnese abdike'ra ˈɔʤʤi 'in faˈvore 'di ˈsuo ˈfiʎʎo |
| 29. | ˈuno ˈʃame 'di 'api invesˈti 'il bamˈbino ˈbjondo kostrin'ʤendolo 'a bbutˈtarsi ˈʤu 'ddall'albero |
| 30. | 'la ferroˈvia 'si znɔˈda ˈlungo 'il ˈfjume segw'ɛndo 'un traʧˈʧato tortuˈoso |
| 31. | ˈdopo aˈvere ˈlɛtto ˈmolti ˈlibri 'luca 'si ri'mise 'a stuˈdjare anˈkora ˈper 'un ˈpɔ |

[3] Although it is often voiced in certain speaking styles and dialects.

32. ˈse arˈrivi ˈal'lalba ˈa k'kapri ˈbutta ˈlankora ˈe pro'segwi ˈa nˈnwɔto
33. ˈʤorʤo 'ʔa ddeˈʧiso ˈdi ˈprɛndere ˈi ˈvoti ˈma ˈprima 'ʔa ddoˈvuto battedzˈdzarsi
34. ˈke ˈne 'faraj ˈdej kwaˈdɛrni ˈdi ˈstɔrja
35. ˈkjɛdi ˈpure ˈa tˈtuo ˈpadre ˈkɔsa 'ne 'pensa dellaŋˈgurja
36. ˈmamma 'e ppa'pa ʻtti 'vɔʎʎono ˈbɛne
37. ˈnon poʤˈʤare 'il bikˈkjɛre ˈkolmo d'akkwa ˈsul pjanoˈfɔrte
38. ˈkon 'la biʧiˈkletta eˈlɛttrika 'le saˈlite ˈsono 'una passeʤˈʤata
39. saˈluta 'la siɲˈɲora 'e ffaj lin'kino
40. 'la teoˈria ˈdej ˈnumeri ˈɛ 'una ˈbranka 'della mateˈmatika
41. ˈmio niˈpote 'ama troˈvare solutsˈtsjoni 'a pproˈblɛmi komˈplɛssi
42. a'gutstsa lin'ʤeɲɲo 'e pro'ʤɛtta 'una ˈradjo intelliˈʤɛnte
43. 'il ʤornaˈlajo ˈvende 'e iɱ'via riˈviste 'e oʤˈʤɛtti tuˈristiʧi
44. 'il ˈkane korˈse forsennataˈmente ˈvɛrso 'il paˈdrone kalpesˈtando ˈle aˈjwɔle
45. 'il doˈlore sor'ʤeva ˈmentre 'la ˈluna 'non ˈɛra anˈkora tramonˈtata
46. pw'ɔi aʧ'ʧɛndere 'la ˈradjo 'a kˈkaso ˈe ssintonidzˈdzarti ˈsu kwalˈsiasi freˈkwɛntsa
47. ˈpɔj ˈʧi ˈsono 'i riˈmɛdi natuˈrali ˈke sˈsono ˈpju effiˈkaʧi 'di ˈtanti proˈdotti preˈsɛnti 'in farmaˈʧia
48. impaˈrjamo 'a mmediˈtare ʤornalˈmente
49. 'si ˈpɛrde koˈsi ˈtanto ˈtɛmpo 'a ddisˈkutere ˈdel ˈnjɛnte
50. staˈsera an'dremo 'al ˈʧinema 'a vveˈdere ˈun ˈfilm franˈʧese
51. ˈsono ˈbɛlli 'i proˈgrammi deˈʧizi 'al'lultimo moˈmento
52. ˈkon krisˈtjana ˈpratiko ˈjɔga ˈoɲɲi merkoleˈdi
53. penˈsjɛri 'e ppaˈrɔle kanˈtava 'la ˈdiva 'kon ˈvoʧe suaˈdɛnte
54. aˈprile 'si esawˈriʃe ˈmentre arˈriva 'kariko 'di proˈmesse 'il 'mese 'di ˈmaʤʤo
55. abˈbjamo trazˈmesso 'il ʤorˈnale ˈradjo 'del matˈtino
56. 'il ˈtɛmpo preˈvisto 'sulli'talja ˈper ˈkwesta ˈsera ˈnon pre'vede temperaˈture 'in awˈmento
57. pen'savo ˈke tˈtu vo'lessi ˈfare ˈsolo ˈuno spunˈtino
58. assikuˈrati 'ke n'non 'si dimentiˈkino 'di ˈskrivere 'alla ˈtsia
59. ˈper salˈvarʧi dob'bjamo resˈtare uˈniti
60. 'il ˈmondo ˈɛ nˈnelle ˈnɔstre ˈmani
61. komporˈtati edukataˈmente 'a tˈtavola
62. ˈpare 'ke sˈsia riˈmasto ˈsolo ˈper 'un ˈkolpo 'di ˈtɛsta
63. 'la ˈpikkola ˈpɛste ˈvwɔle 'il ˈʧuʧʧo ˈper kalˈmarsi
64. 'non ˈmɔrdere 'la ˈspalla 'della ˈnɔnna
65. 'la ˈkarta 'non 'si 'manʤa ˈse 'non ˈsɛj 'una ˈkapra
66. 'il paˈvone bek'ka 'le ˈfɔʎʎe ˈsul viˈale ˈdello ˈdzɔo
67. 'allimprov'vizo 'si uˈdi l'urlo 'del barbaˈʤanni
68. 'basta ˈkon 'i fanaˈtizmi esaʤeˈrati
69. 'non ˈzmettere 'di fantastiˈkare ad ˈɔkki aˈpɛrti
70. ˈkɔl ˈvɛnto ˈin ˈpoppa attraverˈsarono 'il mediterˈraneo 'in 'un ˈsoffjo
71. 'voltati 'e rˈrɛnditi ˈkonto 'di ˈkwanta ˈstrada ˈʔaj perˈkorso
72. 'una ˈtatstsa 'di 'tɛ vˈverde 'al ˈʤorno rinfresˈka 'la ˈmente
73. skrive'rɔ ˈkwesta ˈlɛttera 'kon 'la 'penna 'a sˈsfɛra
74. 'ke 'ne 'pensi 'di 'una ˈfetta 'di ˈtorta
75. 'il ʧesˈtino ˈper 'la ˈkarta ˈsta sˈsotto 'la skrivaˈnia
76. 'una vaˈkantsa 'in agritu'rizmo 'in to'skana 'ʔa 'un ˈkɔsto riˈdotto
77. ˈsul paviˈmento 'del saˈlone 'ʤaʧe 'un tapˈpeto perˈsjano
78. l'albero 'di ˈʧedro ˈɛ sˈsimbolo 'del liˈbano
79. 'un biʎˈʎetto 'di awˈguri akkomˈpaɲɲa 'il reˈgalo
80. tor'neresti 'a kˈkasa 'a pˈpjɛdi
81. 'il ˈgrano saraˈʧɛno 'non kontj'ɛne ˈglutine
82. 'il ˈpane ˈljɛvita ˈkwando 'la ˈluna ˈɛ pˈpjɛna
83. ˈstɛndi 'il buˈkato 'al ˈsole 'e rrisˈparmi enerˈʤia
84. ˈsotto 'la ˈpjatstsa 'ʤaʧe 'un teˈsɔro
85. 'la baˈlena ˈblu 'nwota 'in soliˈtarjo
86. ˈsɛrvono ˈnwɔvi diriˈʤɛnti ˈper rilanˈʧare 'le adzˈdzjɛnde
87. kreˈare laˈvoro ˈɛ 'un doˈvere kostitutstsjoˈnale
88. ˈma ˈkwesta ˈɛ 'un'altra ˈstɔrja 'su kˈkuj 'si indage'ra
89. 'si 'al regolaˈmento 'ke imˈpone ˈlimiti 'alla stupidiˈta
90. 'la balleˈrina in'dossa 'un kosˈtume ˈrɔsa 'fragola
91. law'tore 'si ˈmwɔve 'kon ʃolˈtetstsa ˈnella paˈlude 'delle paˈrɔle
92. 'la faʃˈʃetta 'ʤusta do'vrɛbbe ˈɛssere aljenatsˈtsjone
93. ˈrɛsti 'in kollegaˈmento 'ke rrisponde'ra 'il ˈprimo operaˈtore ˈlibero
94. ve'djamo 'se 'la risˈposta ˈɛ kˈkwella ˈʤusta
95. lavo'kado ˈkreʃʃe ˈnej paˈesi tropiˈkali
96. ˈpeʃʃe ˈfritto 'e insaˈlata ˈmista ˈgratstsje
97. favo'riʃe 'un kafˈfɛ dˈdopo ˈʧena ˈkol diʤesˈtivo
98. 'la ˈfolla ˈɛra impatsˈtsita 'alla ˈvista 'dellassas'sino
99. ˈlɛj ˈkol maʎˈʎone ˈrosso 'si s'stia ˈtsitto
100. manʤeˈrɛbbe volenˈtjɛri 'un fiˈletto 'di bakkaˈla k'kon 'le oˈlive

The LaMIT database includes, in total, 563 words that form the LaMIT lexicon. The list of words and their "computer-friendly" ARPAbet transcription is reported in Appendix 1.

### IV.4 Labeling the LaMIT database

The individual acoustic cues to the features of the words in the LaMIT database were labelled according to the information described in the previous sections. Figure 3 shows an example of a labeled sentence (sentence #36, pronounced by male speaker LDN, first repetition). As shown, a Word tier contains the word segmentation for the sentence under consideration. The segmentation of the sentence into words was done manually by visually examining both the spectrogram and the waveform. Note that word segmentation is always approximate, since cues to preceding and succeeding words often overlap with the temporal extent of the target word. To correctly create the word tier, it was essential to grasp differences among vowels, glides and consonants, as well as the characteristics of the different types of consonants (stops, affricates, fricatives and nasals), in order to correctly associate words with the appropriate region in the waveform. Boundaries were inserted to mark the beginning and end of each word. All 800 sentences of the database were manually labeled in this way.

Figure 3 also shows the predicted LEXI phoneme tier. This tier was automatically generated by a computer program that reads from the Word tier, matches each word with its entry in the lexicon (see transcription in Appendix A), and uses the phonological specification in the lexical entry to output the predicted sequence of phonemes. As shown in Figure 3, the automatic program that generates the LEXI tier subdivides the word interval into phoneme intervals, and assigns a phoneme to each interval. Here we would like to make the same point as about words above, i.e. that there are no boundaries between phonemes in the signal. We use landmarks to estimate the onset and offset of words and phonemes, despite the fact that there is extensive overlap of cues. These linguistic constituents exist only in the mind of the speaker and the listener (and not in the signal). However, landmarks approximate the boundaries in a convenient way. The actual onset and offset of each phoneme are then manually adjusted, in a following labelling phase, as shown in Fig. 4 for the same example sentence.

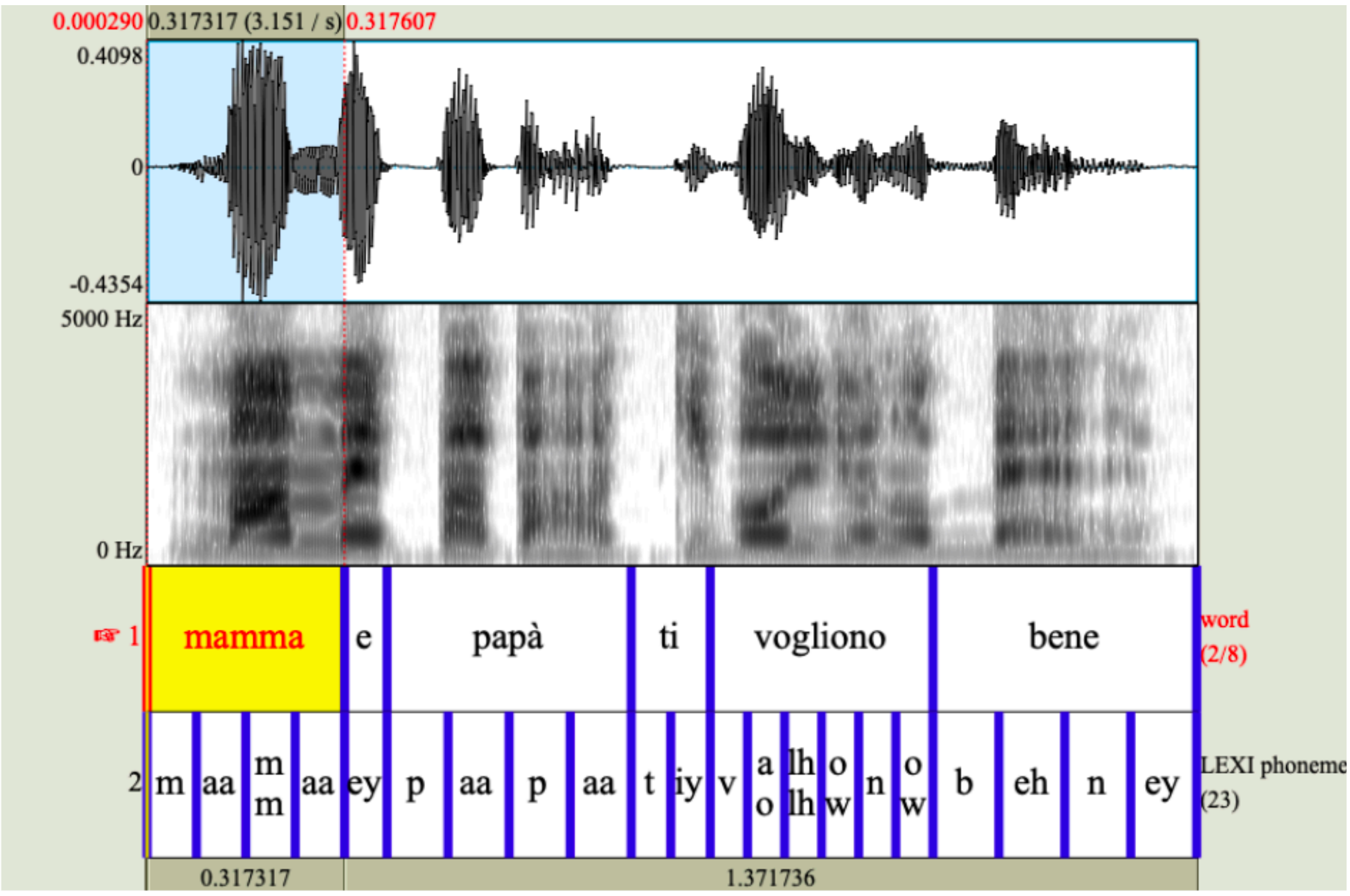

Figure 3 – Labeling the LaMIT sentences. The figure shows the example of sentence #36 (see Tables VII and VIII) produced by male speaker LDN (first repetition), and in particular shows the waveform, spectrogram, word tier, and predicted phoneme tier (called the LEXI phoneme tier). This predicted phoneme tier is generated automatically, based on the LaMIT lexicon (see Appendix A): phonemic boundaries are manually adjusted as shown on Fig. 4.

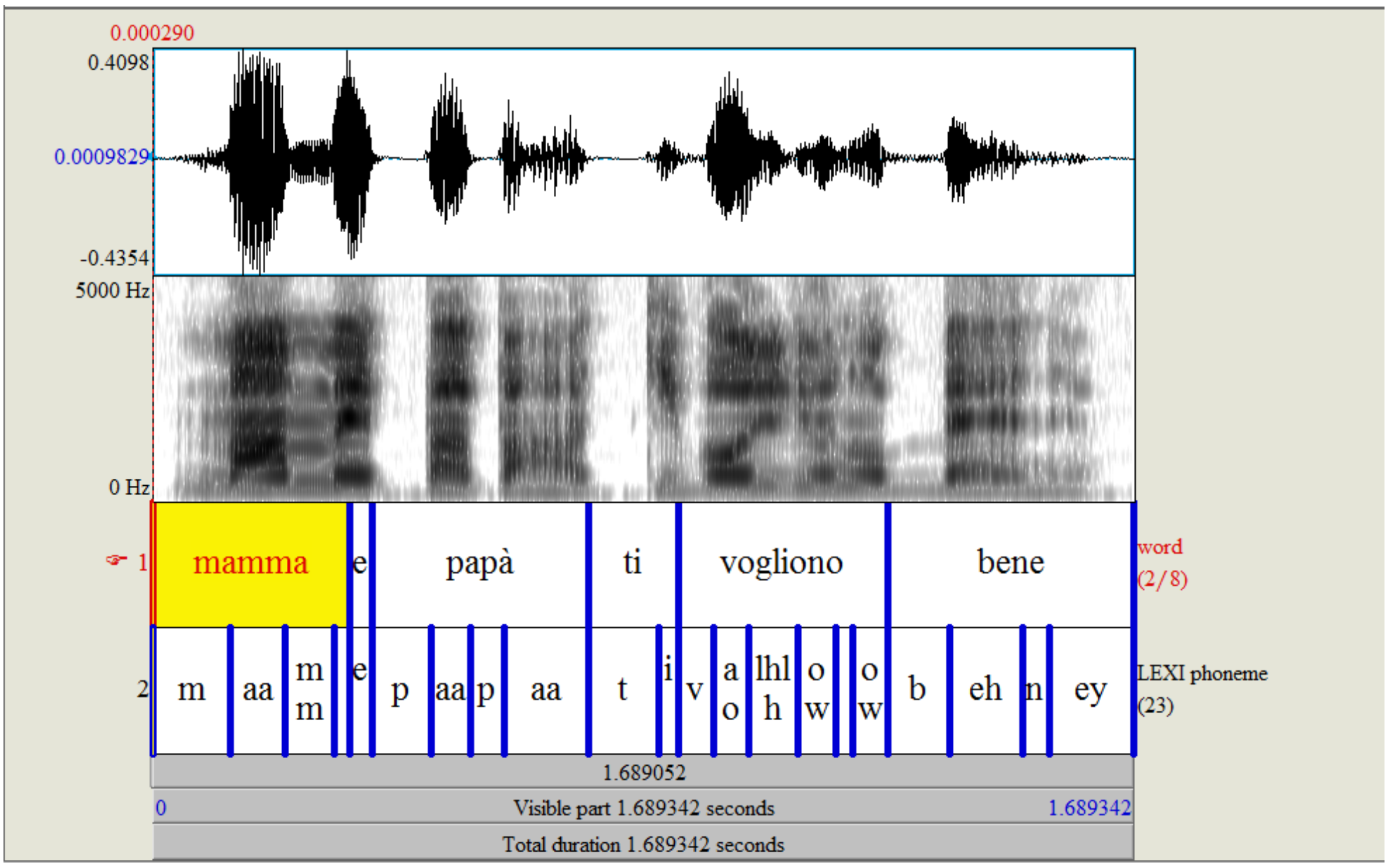

Figure 4 – Example of manually adjusted boundaries for the phonemes of sentence #36, by male speaker LDN, first repetition (compare with not-yet-adjusted labels for the same sentence in Fig. 3).

The next step in the labeling process is to generate a modified LEXI phoneme tier, according to the actual manually detected and perceived phonemes in the sentence. For example, if a speaker pronounces an intervocalic /s/ as /z/, as most common in some Italian dialects, the modified LEXI tier will annotate /z/ rather than the predicted/s/.

Additional tiers contain information on relevant instants of time for the detection of cues to articulator-free and articulator-bound features, as listed in Section III. These include landmarks that identify abrupt acoustic changes associated with consonant closures and releases, as well as regions of maximum opening of the vocal tract, signaling vowels landmarks, annotated in the Landmark tier, and its associated modified Landmark ties. The "modified" tier contains, as for the modified LEXI tier, those landmarks that were actually manually detected in the sentence. Related to articulator-bound features, stand:

→ The Glottal tier: used to mark glottal sounds that originate from the manipulation of the air flow from the vocal tract (glottis)

→ The Nasal tier: used to mark nasal intervals

→ The Cplace tier: used to mark consonant intervals, i.e. formant transition closure, initial sound release, and formant transition release

→ The Vgplace tier: used to indicate properties of vowels, i.e. height, roundness, and locality.

Figure 5 shows the same sentence in Figs. 3 and 4, fully labeled; Figure 6 zooms on the first word of Fig. 5 to better show the different labels in the different tiers.

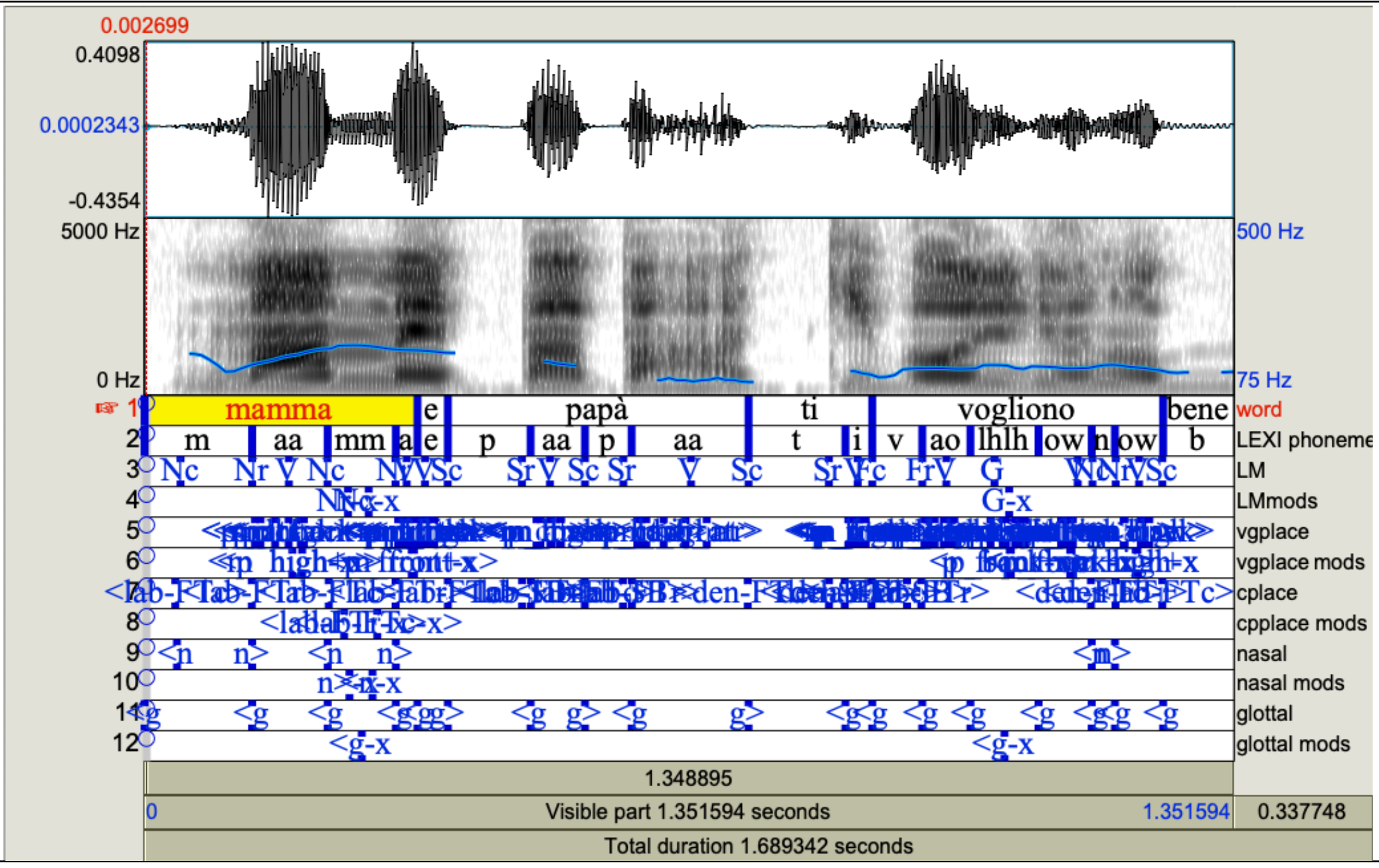

Figure 5 – Same sentence as in Figs. 3 and 4, fully labeled. Note the presence of the Landmark tier, the Cplace tier, the Vgplace tier, the Nasal tier, the Glottal tier, and their modified counterparts.

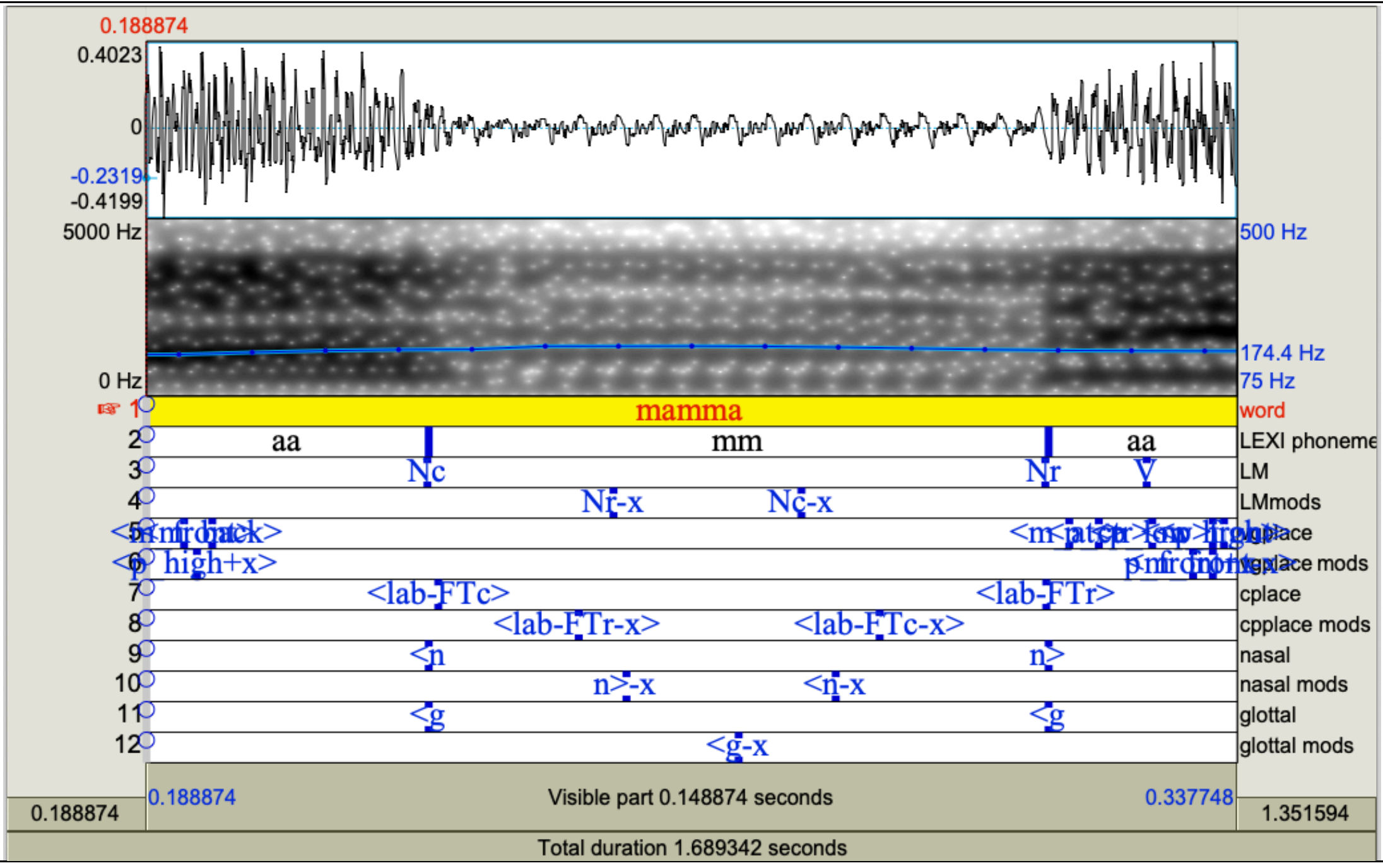

Figure 6 – Zoom on the first word of the sentence shown in Figures 3-5, fully labeled.

## V. Proposed extension of the model and future research directions

A first and fundamental question that may be addressed is: which parts of the system are suitable for the introduction of the concept of *inference* – an absent concept in Stevens' model – and for what reasons? How does the introduction of inference impact the decision rules that lead to the recognition of words? We postulate that the physical layer of the system, that is that low-level processing of a speech signal toward the extraction of acoustic cues, is not statistical in nature. That is, the mechanisms that detect acoustic cues to distinctive features do not attempt to infer whether partial or incomplete information signals a feature or not, and therefore do not take any particular action when an acoustic cue – a landmark for example – is too weak to be detected or absent. In other words, these mechanisms do not attempt to estimate the probability of an event associated with a variable degree of physical evidence. To better explain this hypothesis, imagine the physical system at hand, i.e. the speech production system of a given speaker, producing multiple instances of the same word, under the same experimental conditions. No matter how similar these conditions may be, it is well known that no utterance can be perfectly duplicated, and therefore all instances will be different, although they all represent that same word intended by the speaker. In other words, the physically-produced acoustic signals corresponding to the multiple instances, although all different, carry information that is relevant to the listener, by adequately representing the abstraction of the word. That abstract representation is, in our view, the word's description in terms of bundles of distinctive features. Suppose that now we manipulate these instances to produce what we call a "prototype" that can be then matched to the abstraction, by for example creating an average pattern from all the instances at hand. We know that the system under examination, a human speaker, will almost never be able to match this prototype. Modeling the system based on an average pattern, or any other pattern that attempts to merge the different instances, is an ill-based approach, since it attempts to model not the physical system itself, which will never produce the artificial pattern, but rather an artificial version of it. The information that all instances carry and that unifies them in a family representing the abstraction is what we define as cues to features. As suggested by Lindau (1985) in a study on /r/, there may be no unique physical property that group instances in the class of rhotics, but rather some kind of family resemblance (Wittgenstein, 1958) according to which pairs of elements in the class may resemble each other on one aspect, while other pairs on other aspects, although all members end up being connected within a certain number of degrees of separation.

Although the acoustic signal may contain all necessary cues to a given feature, it must transit through the auditory system from the basilar membrane up to the cortex, where it elicits the neural response that is then processed by the brain. The transport of information from one neuron to the next is achieved by neurotransmitters, a kind of biochemical messenger (Friederici, 2017). The presence of some kind of "neuronal noise" may require statistical modelling to allow correct decoding and denoising under adverse neuronal noise conditions and to provide a model for robust cues detection in the brain. On another front stands the system that, from the set of possible lexical candidates derived from the extracted cues, takes a decision in favor of one entry from the cohort of candidate words, and by doing so completes the lexical access process. The cohort model (Marslen-Wilson and Welsh, 1978) suggests that this process may be sequential, shrinking the cohort as one goes through the sequence of phonemes. There seems to be at least one reason to propose that this system may operate according to statistical principles: in estimating the "best" candidate, the system must function according to a best hypothesis principle, following a Bayesian approach for example, and therefore perform a statistical estimation of what is the best candidate, based on a pre-defined similarity metric that should quantify the distance of candidates to reference patterns. There is a large literature on lexical similarity (see for example Andrews, 1992; Islam and Inkpen, 2008), but to our knowledge no study addresses the problem of defining a similarity metric that accounts for a hierarchical organization of distinctive features, so that some features count more than others. The question of whether the hierarchical organization of Stevens' lexical access model reflecting articulatory gestures and a hierarchy of distinctive features is preserved when it comes to defining the perceptual distance between two lexical hypotheses, and of how to design a distance metric that infers a "best" lexical choice based on a hierarchically organized inventory of distinctive features, should be addressed in future work. Figure 7 shows a possible expansion of the Lexical Access schematic representation presented in Fig.1, to illustrate some proposed future research directions.

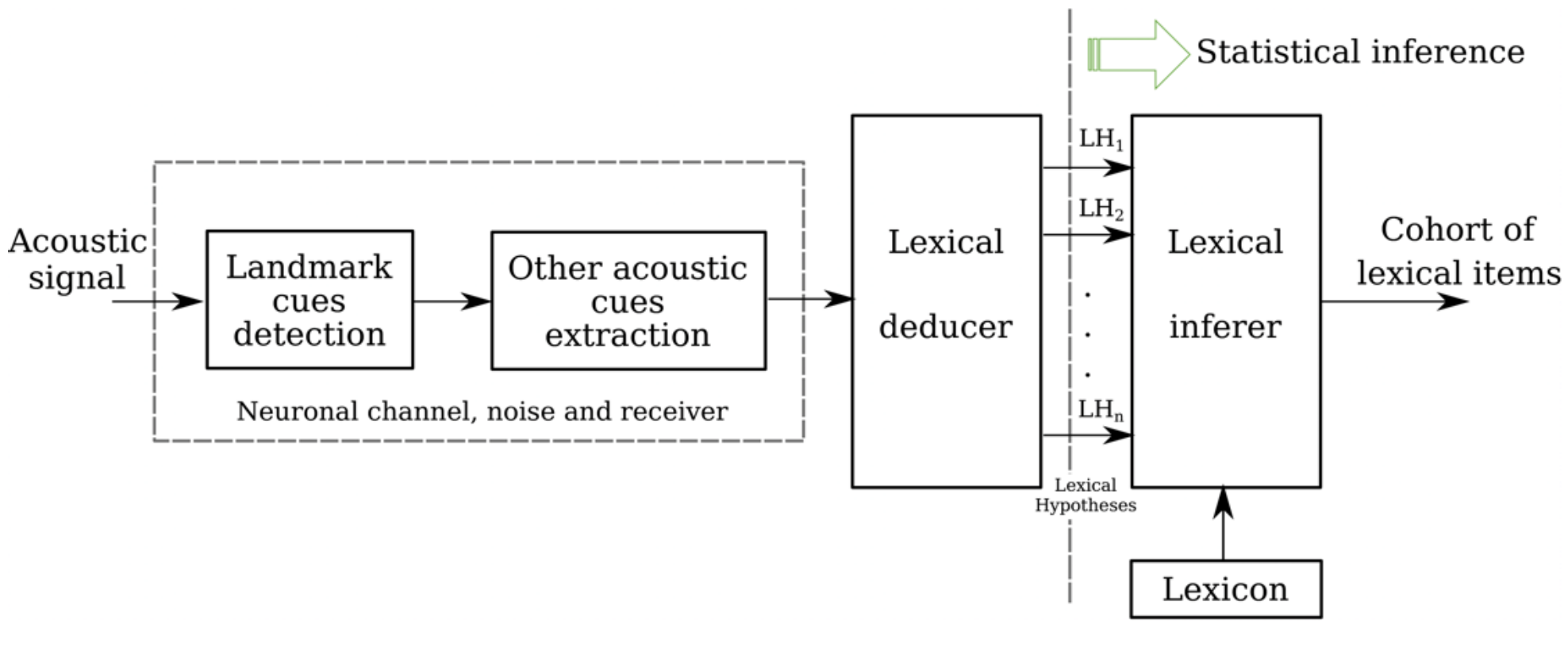

Figure 7 – Proposed extension of the Lexical Access system

## VI. Conclusion

This paper has presented a model and framework for the design of a speech recognition system for the Italian language, based on Stevens' model of Lexical Access (Stevens, 2002). After providing a synthesis of the founding principles of Stevens' model, the paper described the Lexical Access system developed for American English in the Speech Communication Group at the Massachusetts Institute of Technology (MIT). The hypotheses that were needed in order to apply the model to Italian were then discussed, addressing, in particular, the phonemic classification that was adopted, including adjustments that occur across word boundaries (such as syntactic doubling - a typical effect of Italian), the LaMIT corpus that was created accordingly, and the labeling process. The LaMIT corpus and labeling information is freely accessible (address: http://acts.ing.uniroma1.it/project_lamit_database.php) under an open-source Creative Commons license.

The forthcoming expected outcomes are:

1. The Lexical Access model has so far only been applied to American-English. Its application to a different language may lead to an understanding of the underlying universal language-independent aspects of the model. In particular, the model postulates the relevance of specific acoustic discontinuities - the landmarks - that correspond to categorized distinctive features (i.e. manner features). Are landmarks language-independent?
2. The development of an Italian speech recognizer based on detection of landmarks and other acoustic cues to distinctive features, and the possibility to provide open access to data and algorithms with the creation of a publicly accessible website, may lead to establishing a reference record for the speech community in Italy and abroad, based on the impact the research may provide in terms of supporting evidence for universal processes of speech perception.

Finally, future research directions were identified. In particular, understanding the role of inference in human speech perception may lead to a major advance in terms of principled modeling of lexical access and speech recognition systems. Moreover, it was suggested that modeling the effect of the transit of the information signal through the auditory system, from the basilar membrane up to the cortex, where it undergoes further neurocognitive process, may lead to cutting-edge contributions and pave the way toward the development of a conceptual model of neuronal processing of acoustic cues. The development and expansion of the lexical access system, based on the landmark theory that is described in this paper, is the fundamental and founding step toward the long-term goal of exploring how cues to features manifest in the brain, by conceiving and carrying out specific neurophysiological experiments - thanks to collaborative research possibly ensuing from this work - that would lead to the consolidation of Stevens' model and of our understanding of how humans understand speech.

# Appendix A

The LaMIT lexicon and ARPAbet transcription of words in the lexicon

```
A AA1
ABBIAMO AA BB Y AA1 M OW
ABDICHERÀ AA B D IY K EY R AA1
ACCADEMICO AA KK AA D EH1 M IY K OW
ACCENDERE AA CHCH EH1 N D EY R EY
ACCOMPAGNA AA KK OW M P AA1 GNGN AA
ACQUISTARLO AA KK W IY1 S T AA R L OW
AD AA1 D
ADDIO AA DD IY1 OW
AGLI AA1 LHLH IY
AGRITURISMO AA G R IY T UW R IY1 Z M OW
AGUZZA AA G UW1 TSTS AA
AIUOLE AA Y W AO1 L EY
AIUTANO AA Y UW1 T AA N OW
AL AA1 L
ALCUNI AA L K UW1 N IY
ALIENAZIONE AA L Y EY N AA TSTS Y OW1 N EY
ALL’ALBA AA LL AA1 L B AA
ALL’IMBRUNIRE AA LL IY M B R UW N IY1 R EY
ALL’IMPERATORE AA LL IY M P EY R AA T OW1 R EY
ALL’IMPROVVISO AA LL IY M P R OW VV IY1 S OW
ALL’ULTIMO AA LL UW1 L T IY M OW
ALLA AA1 LL AA
ALLE AA1 LL EY
ALLENAMENTO AA LL EY N AA M EY1 N T OW
ALZARTI AA L TS AA1 R T IY
AMA AA1 M AA
ANCORA AA N K OW1 R AA
ANDREMO AA N D R EY1 M OW
APERTI AA P EY1 R T IY
API AA1 P IY
APPONI AA PP OW1 N IY
APPUNTAMENTO AA PP UW N T AA M EY1 N T OW
APRILE AA P R IY1 L EY
ARRIVA AA RR IY1 V AA
ARRIVI AA RR IY1 V IY
ASSICURATI AA SS IY K UW R AA1 T IY
ATEO AA1 T EY OW
ATTENERSI AA TT EY N EY1 R S IY
ATTRAVERSARONO AA TT R AA V EY R S AA1 R OW N OW
AUGURI AA W G UW1 R IY
AUMENTO AA W M EY1 N T OW
AVERE AA V EY1 R EY
AVER AA V EY1 R
AVVENTURA AA VV EY N T UW1 R AA
AZIENDE AA DZDZ Y EH1 N D EY
BACCALÀ B AA KK AA L AA1
BALENA B AA L EY1 N AA
BALLERINA B AA LL EY R IY1 N AA
BAMBINO B AA M B IY1 N OW
BANCA B AA1 N K AA
BARBAGIANNI B AA R B AA JH AA1 NN IY
BASSO B AA1 SS OW
BASTA B AA1 S T AA
BATTEZZARSI B AA TT EY DZDZ AA R S IY
BEBÈ B EY B EH1
BECCA B EY1 KK AA
BELLI B EH1 LL IY
BENE B EH1 N EY
BIANCO B Y AA1 N K OW
BIBLIOTECA B IY B L Y OW T EH1 K AA
BICCHIERE B IY KK Y EH1 R EY
BICICLETTA B IY CH IY K L EY1 TT AA
BIGLIETTO B IY LHLH EY1 TT OW
BIONDO B Y OW1 N D OW
BLU B L UW1
BRANCA B R AA1 N K AA
BUCATO B UW K AA1 T OW
BUTTA B UW1 TT AA
BUTTARSI B UW TT AA1 R S IY
CAFFÈ K AA FF EH1
CALCE K AA1 L CH EY
CALMARSI K AA L M AA1 R S IY
CALPESTANDO K AA L P EY S T AA1 N D OW
CANE K AA1 N EY
CANTAVA K AA N T AA1 V AA
CAPRA K AA1 P R AA
CAPRI K AA1 P R IY
CARICO K AA1 R IY K OW
CARTA K AA1 R T AA
CARTOLINE K AA R T OW L IY1 N EY
CASA K AA1 S AA
CASO K AA1 S OW
CASSETTO K AA SS EY1 TT OW
CATTEDRALE K AA TT EY D R AA1 L EY
CEDRO CH EY1 D R OW
CELLULARE CH EY LL UW L AA1 R EY
CENA CH EY1 N AA
CESTINO CH EY S T IY1 N OW
CHE K EY1
CHIEDI K Y EH1 D IY
CHIUDERE K Y UW1 D EY R EY
CI CH IY1
CIELO CH EH1 L OW
CINEMA CH IY1 N EY M AA
CIOCCOLATO CH OW KK OW L AA1 T OW
CITTÀ CH IY TT AA1
CIUCCIO CH UW1 CHCH OW
CLIENTI K L IY EH1 N T IY
COL K AO1 L
COLLEGAMENTO K OW LL EY G AA M EY1 N T OW
COLMO K OW1 L M OW
COLPO K OW1 L P OW
COMPLESSI K OW M P L EH1 SS IY
COMPORTATI K OW M P OW1 R T AA T IY
COMPRA K OW1 M P R AA
CON K OW1 N
CONTIENE K OW N T Y EH1 N EY
CONTO K OW1 N T OW
CONTRATTO K OW N T R AA1 TT
CORRENTE K OW RR EH1 N T EY
CORSE K OW1 R S EH
COSA K AO1 S AA
COSÌ K OW S IY1
COSTITUZIONALE K OW S T IY T UW TSTS Y OW N AA1
L EY
COSTO K AO1 S T OW
COSTRINGENDOLO K OW S T R IY N JH EY1 N D OW L
OW
COSTUME K OW S T UW1 M EY
CREARE K R EY AA1 R EY
CRESCE K R EY1 SHSH EY
CRISTIANA K R IY S T Y AA1 N AA
CUGINO K UW JH IY1 N OW
CUI K UW1 Y
CUORE K W AO1 R EY
D’ACQUA D AA1 KK W AA
D’ACCORDO D AA KK AO1 R D OW
D’AMORE D AA M OW1 R EY
DA D AA1
DAL D AA1 L
DALL’ALBERO D AA LL AA1 L B EY R OW
DECIDERTI D EY CH IY1 D EY R T IY
DECISI D EY CH IY1 S IY
DECISO D EY CH IY1 S OW
DEI D EY1 Y
DEL D EY1 L
DELL’ANGURIA D EY LL AA N G UW1 R Y AA
DELL’ANNO D EY LL AA1 NN OW
DELL’ASSASSINO D EY LL AA SS AA SS IY1 N OW
DELL’INVERNO D EY LL IY N V EH1 R N OW
DELLA D EY1 LL AA
DELLE D EY1 LL EY
DELLO D EY1 LL OW
DEVONO D EH1 V OW N OWDI D IY1
DIGESTIVO D IY JH EY S T IY1 V OW
DIGITALE D IY JH IY T AA1 L EY
DIMENTICHINO D IY M EY1 N T IY K IY N OW
DIRIGENTI D IY R IY JH EH1 N T IY
DIROTTO D IY R OW1 TT OW
DISCENDI D IY SHSH EY N D IY
DISCUTERE D IY S K UW1 T EY R EY
DISSE D IY1 SS EY
DIVA D IY1 V AA
DOBBIAMO D OW BB Y AA1 M OW
DOLORE D OW L OW1 R EY
DOPO D OW1 P OW
DOVER D OW V EY1 R
DOVERE D OW V EY1 R EY
DOVRAI D OW V R AA1 Y
DOVREBBE D OW V R EY1 BB EY
DOVUTO D OW V UW1 T OW
DUE D UW1 EY
E EY1
È EH1
EDUCATAMENTE EY D UW K AA T AA M EY1 N T EY
EFFICACI EY FF IY K AA1 CH IY
ELETTRICA EY L EH1 TT R IY K AA
ENERGIA EY N EY R JH IY1 AA
ENIGMISTICA EY N IY G M IY1 S T IY K AA
ERA EH1 R AA
ESAGERATI EY S AA JH EY R AA1 T IY
ESAURISCE EY S AA W R IY1 SH EY
ESSERE EH1 SS EY R EY
FAI F AA1 Y
FAMOSO F AA M OW1 S OW
FANATISMI F AA N AA T IY1 Z M IY
FANTASIA F AA N T AA S IY1 AA
FANTASTICARE F AA N T AA S T IY K AA1 R EY
FARAI F AA R AA1 Y
FARE F AA1 R EY
FARMACIA F AA R M AA CH IY1 AA
FASCETTA F AA SHSH EY1 TT AA
FAVORE F AA V OW1 R EY
FAVORISCE F AA V OW R IY SH EY
FERROVIA F EY RR OW V IY1 AA
FETTA F EY1 TT AA
FIAMME F Y AA1 MM EY
FIGLIO F IY1 LHLH OW
FILETTO F IY L EY1 TT OW
FILM F IY1 L M
FIRMA F IY1 R M AA
FISSERANNO F IY SS EY R AA1 NN OW
FIUME F Y UW1 M EY
FOGLIE F AO1 LHLH EY1
FOLLA F AO1 LL AA
FORSENNATAMENTE F OW R S EY NN AA T AA M EY1 N T
EY
FORTE F AO1 R T EY
FOTOGRAFIA F OW T OW G R AA F IY1 AA
FRAGOLA F R AA1 G OW L AA
FRANCESE F R AA N CH EY1 S EY
FREQUENZA F R EY K W EH1 N TS AA
FRITTO F R IY1 TT OW
FUGGE F UW JHJH EY
FURONO F UW1 R OW N OW
GAMBE G AA1 M B EY
GATTO G AA1 TT OW
GATTONARE G AA TT OW N AA1 R EY
GIACE JH AA1 CH EY
GIAPPONESE JH AA PP OW N EY1 S EY
GIARDINO JH AA R D IY1 N OW
GIORGIO JH OW1 R JH OW
GIORNALAIO JH OW R N AA L AA1 Y OW
GIORNALE JH OW R N AA1 L EY
GIORNALMENTE JH OW R N AA L M EY1 N T EY
GIORNO JH OW1 R N OW
GIÙ JH UW1
GIUSTA JH UW1 S T AA
GLADIOLI G L AA D IY1 OW L IY
GLI LH IY1
GLUTINE G L UW1 T IY N EY
GNOMI GN OW1 M IY
GODI G AO1 D IY
GOVERNO G OW V EH1 R N OW
GRAFFIATO G R AA FF Y AA1 T OW
GRANO G R AA1 N OW
GRAZIE G R AA1 TSTS Y EY
GUARDARE G W AA R D AA1 R EY
HA AA1
HAI AA1 Y
I IY1
IL IY1 L
IMMAGINI IY MM AA1 JH IY N IY
IMPARIAMO IY M P AA R Y AA1 M OW
IMPAZZITA IY M P AA TSTS IY1 T AA
IMPONE IY M P OW1 N EY
IN IY1 N
INDAGHERÀ IY N D AA G EY R AA1
INDIMENTICABILE IY N D IY M EY N T IY K
AA1 B IY L EY
INDOSSA IY N D OW1 SS AA
INQUISITORI IY N K W IY S IY T OW1 R IY
INSALATA IY N S AA L AA1 T AA
INTELLIGENTE IY N T EY LL IY JH EH1 N T EY
INVESTÌ IY N V EY S T IY1
INVIA IY N V IY1 AA
ITALIANA IY T AA L Y AA1 N AA
L’INCHINO L IY N K IY1 N OW
L’INGEGNO L IY N JH EY1 GNGN OW
L’ALBERO L AA1 L B EY1 R OW
L’ANCORA L AA1 N K OW R AA
L’AUTORE L AA W T OW1 R EY
L’AVOCADO L AA V OW K AA1 D OW
L’IMPERATORE L IY M P EY R AA T OW1 R EY
L’UNIVERSITÀ L UW N IY V EY R S IY T
AA1L’URLO L UW1 R L OW
LA L AA1
LADRO L AA1 D R OW
LAUREARSI L AA W R EY AA1 R S IY
LAVORO L AA V OW1 R OW1
LE L EY1
LEGGERE L EH1 JHJH EY R EY
LEI L EH1 Y
LETTERA L EY1 TT EY R AA
LETTO L EH1 TT OW
LEVATE L EY V AA1 T EY
LIBANO L IY B AA1 N OW
LIBERO L IY1 B EY R OW LIBRI L IY1 B R IY
LIBRO L IY1 B R OW
LIEVITA L Y EH1 V IY T AA
LIMITI L IY1 M IY T IY
LO L OW1
LUCA L UW1 K AA
LUGLIO L UW1 LHLH OW
LUNA L UW1 N AA
LUNGO L UW1 N G OW
MA M AA1
MACHIAVELLI M AA K Y AA V EY1 LL IY
MAGGIO M AA1 JHJH OW
MAGLIONE M AA LHLH OW1 N EY
MALGRADO M AA L G R AA1 D OW
MAMMA M AA1 MM AA
MANDARE M AA N D AA1 R EY
MANGEREBBE M AA N JH EY R EH1 BB EY
MANGIA M AA1 N JH AA
MANI M AA1 N IY
MARIA M AA R IY1 AA
MATEMATICA M AA T EY M AA1 T IY K AA
MATTINO M AA TT IY1 N OW
MEDITARE M EY D IY T AA1 R EY
MEDITERRANEO M EY D IY T EY RR AA1 N EY OW
MENO M EY1 N OW
MENTE M EY1 N T EY
MENTRE M EY1 N T R EY
MERAVIGLIOSA M EY R AA V IY LHLH OW1 S AA
MERCOLEDÌ M EY R K OW L EY D IY1
MESE M EY1 S EY
MIO M IY1 OW
MISTA M IY1 S T AA
MODA M AO1 D AA
MOLTI M OW1 L T IY
MOMENTO M OW M EY1 N T OW
MONDO M OW1 N D OW
MONTATA M OW N T AA1 T AA
MONTE M OW1 N T EY
MORDERE M AO1 R D EY R EY
MUOVE M W AO1 V EY
MUSICA M UW1 S IY K AA
NATURALI N AA T UW R AA1 L IY
NE N EY1
NECESSARIO N EY CH EY SS AA1 R Y OW
NEI N EY1 Y
NEL N EY1 L
NELLA N EY1 LL AA
NELLE N EY1 LL EY
NICCOLÒ N IY KK OW L OW1
NIENTE N Y EH1 N T EY
NIPOTE N IY P OW1 T EY
NON N OW1 N
NONNA N AO1 NN AA
NOSTALGIA N OW S T AA L JH IY1 AA
NOSTRE N AO1 S T R EY
NUDI N UW1 D IY
NUMERI N UW1 M EY R IY
NUOTA N W OW1 T AA
NUOTO N W AO1 T OW
NUOVI N W AO1 V IY
O AO1
OCCHI AO1 KK IY
```

The LaMIT lexicon and ARPAbet transcription of words in the lexicon (continued)

OGGETTI OW JHJH EH1 TT IY
OGGI AO1 JHJH IY
OGNI OW1 GNGN IY
OLIVE OW L IY1 V EY
OPACIZZATO OW P AA CH IY DZDZ AA1 T OW
OPERATORE OW P EY R AA T OW1 R EY
OPERE AO1 P EY R EY
PADRE P AA1 D R EY
PADRONE P AA1 D R OW N EY
PAESI P AA EY1 S IY
PALUDE P AA L UW1 D EY
PANE P AA1 N EY
PANNA P AA1 NN AA
PAPÀ P AA P AA1
PARE P AA1 R EY
PAROLE P AA R AO1 L EY
PASSA P AA1 SS AA
PASSEGGEREI P AA SS EY JHJH EH R EH1 Y
PASSEGGIATA P AA SS EY JHJH AA1 T AA
PAVIMENTO P AA V IY M EY1 N T OW
PAVONE P AA V OW1 N EY
PAZZO P AA1 TSTS OW
PELOSO P EY L OW1 S OW
PENNA P EY1 NN AA
PENSA P EY1 N S AA
PENSAVO P EY1 N S AA1 V OW
PENSERESTI P EY N S EY R EY1 S T IY
PENSI P EY1 N S IY
PENSIERI P EY N S Y EH1 R IY
PENSÒ P EY N S AO1
PER P EY1 R
PERCHÉ P EY R K EY1
PERCORSO P EY R K OW1 R S OW
PERDE P EH1 R D EY
PERSIANO P EY R S Y AA1 N OW
PESCE P EY1 SHSH EY
PESTE P EH1 S T EY
PIANOFORTE P Y AA N OW F AO1 R T EY
PIAZZA P Y AA1 TSTS AA
PICCOLA P IY1 KK OW L AA
PIEDI P Y EH1 D IY
PIENA P Y EH1 N AA
PIENO P Y EH1 N OW
PIETRO P Y EH1 T R OW
PINZA P IY1 N TS AA
PIOVE P Y AO1 V EY
PIÙ P Y UW1
PO' P AO1
POETA P OW EH1 T AA
POGGIARE P OW JHJH AA1 R EY
POI P AO1 Y
POPPA P OW1 PP AA
POTENDO P OW T EH1 N D OW
POTRANNO P OW T R AA1 NN OW
PRATICO P R AA1 T IY K OW
PRECARIA P R EY K AA1 R Y AA
PRENDERE P R EH1 N D EY R EY
PRENDI P R EH1 N D IY
PRESENTI P R EY S EH1 N T IY
PRESTITO P R EH1 S T IY T OW
PRESTO P R EH1 S T OW
PREVEDE P R EY V EY1 D EY
PREVISTO P R EY V IY1 S T OW
PRIMA P R IY1 M AA
PRIMO P R IY1 M OW
PRIVARSI P R IY V AA1 R S IY
PROBLEMI P R OW B L EH1 M IY
PRODOTTI P R OW D OW1 TT IY
PROGETTA P R OW JH EY1 TT AA
PROGRAMMI P R OW G R AA1 MM IY
PROMESSE P R OW M EY1 SS EY
PROSEGUI P R OW S EY1 G W IY
PUBBLICA P UW1 BB L IY K AA
PUOI P W AO1 IY
PUR P UW1 R
PURE P UW1 R EY
QUADERNI K W AA D EH1 R N IY
QUALSIASI K W AA L S IY1 AA S IY
QUANDO K W AA1 N D OW
QUANTA K W AA1 N T AA
QUELLA K W EY1 LL AA
QUESTA K W EY1 S T AA
RADIO R AA1 D Y OW
RAGAZZI R AA G AA1 TSTS IY
RAPPRESENTAZIONE R AA PP R EY S EY N T AA TSTS Y OW1 N EY
REGALO R EY G AA1 L OW
REGOLAMENTO R EY G OW L AA M EY1 N T OW
REGOLE R EH1 G OW L EY
RELIGIOSA R EY L IY JH OW1 S AA
REMOTE R EY M AO1 T EY
RENDERE R EH1 N D EY R EY
RENDITI R EH1 N D IY T IY
RESSE R EH1 SS EY
RESTARE R EY S T AA1 R EY
RESTI R EH1 S T IY
RIDOTTO R IY D OW1 TT OW
RILANCIARE R IY L AA N CH AA1 R EY
RIMASTO R IY M AA1 S T OW
RIMEDI R IY M EH1 D IY
RIMISE R IY M IY1 S EY
RINFORZA R IY N F AO1 R TS AA
RINFRESCA R IY N F R EY1 S K AA
RIPORRE R IY P OW1 RR EY
RISPARMI R IY S P AA1 R M IY
RISPONDERÀ R IY S P OW N D EY R AA1
RISPOSTA R IY S P OW1 S T AA
RIVISTE R IY V IY1 S T EY
RIVOLGENDOSI R IY V AO L JH EY1 N D OW S IY
RIVOLUZIONATO R IY V OW L UW TSTS Y OW N AA1 T OW
ROSA R AO1 S AA
ROSSO R OW1 SS OW
SALITE S AA L IY1 T EY
SALONE S AA L OW1 N EY
SALUTA S AA L UW1 T AA
SALUTARE S AA L UW T AA1 R EY
SALVARCI S AA L V AA1 R CH IY
SARACENO S AA R AA CH EH1 N OW
SCAPPA S K AA1 PP AA
SCHERMO S K EY1 R M OW SCI SH IY1
SCIAME SH AA1 M EY
SCIENZIATO SH EY N TS Y AA1 T OW
SCIOLTEZZA SH OW L T EY1 TSTS AA
SCRIVANIA S K R IY V AA N IY1 AA
SCRIVERE S K R IY1 V EY R EY
SCRIVERÒ S K R IY V EY R AO1
SCUOLA S K W AO1 L AA
SE S EY1
SEGUENDO S EY G W EH1 N D OW
SEI S EH1 Y
SENTI S EY1 N T IY
SERA S EY1 R AA
SERVONO S EH1 R V OW N OW
SETTIMANA S EY TT IY M AA1 N AA
SFERA S F EH1 R AA
SI S IY1
SIA S IY1 AA
SIGNORA S IY GNGN OW1 R AA
SIMBOLO S IY1 M B OW L OW
SINDACO S IY1 N D AA K OW
SINTONIZZARTI S IY N T OW N IY DZDZ AA1 R T IY
SMETTERE Z M EY1 TT EY R EY
SNODA Z N OW1 D AA
SOFFIO S OW1 FF Y OW
SOGNI S OW1 GNGN IY
SOLE S OW1 L EY
SOLITARIO S OW L IY T AA1 R Y OW
SOLO S OW1 L OW
SOLUZIONI S OW L UW TSTS Y OW1 N IY
SONO S AO1 N OW
SORELLE S OW R EH1 LL EY
SORGEVA S OW R JH EY1 V AA
SOTTO S OW1 TT OW
SPALLA S P AA1 LL AA
SPUNTINO S P UW N T IY1 N OW
STA S T AA1
STABILITE S T AA B IY L IY1 T EY
STASERA S T AA S EY1 R AA
STATO S T AA1 T OW
STENDI S T EH1 N D IY
STIA S T Y AA1
STORIA S T AO1 R Y AA
STRADA S T R AA1 D AA
STRUTTURA S T R UW TT UW1 R AA
STUDENTI S T UW D EH1 N T IY
STUDIARE S T UW D Y AA1 R EY
STUPIDITÀ S T UW P IY D IY T AA1
SU S UW1
SUADENTE S UW AA D EH1 N T EY
SUFFICIENTI S UW FF IY CH EH1 N T IY
SUL S UW1 L
SULL'ITALIA S UW LL IY T AA1 L Y AA
SUO S UW1 OW
SVETTA Z V EY1 TT AA
TANTI T AA1 N T IY
TANTO T AA1 N T OW
TAPPETO T AA PP EY1 T OW
TAVOLA T AA1 V OW L AA
TAZZA T AA1 TSTS AA
TÈ T EH1
TEMPERATURE T EY M P EY R AA T UW1 R EY
TEMPO T EH1 M P OW
TEORIA T EY OW R IY1 AA
TESORO T EY S AO1 R OW
TESTA T EH1 S T AA
TI T IY1
TORNERESTI T OW R N EY1 R EY S T IY
TORTA T OW1 R T AA
TORTUOSO T OW R T UW OW1 S OW
TOSCANA T OW S K AA1 N AA
TRACCIATO T R AA CHCH AA1 T OW
TRAMONTATA T R AA M OW N T AA1 T AA
TRANSAZIONE T R AA N S AA TSTS Y OW1 N EY
TRASMESSO T R AA Z M EY1 SS OW
TROPICALI T R OW P IY K AA1 L IY
TROVARE T R OW V AA1 R EY
TU T UW1
TUO T UW1 OW
TURISTICI T UW R IY1 S T IY1 CH IY
UDÌ UW D IY1
UGGIOSO UW JHJH OW1 S OW
UN UW1 N
UN'ALTRA UW N AA1 L T R AA
UN'ESPERIENZA UW N EY S P EY R Y EH1 N TS AA
UN'ISTITUZIONE UW N IY S T IY T UW TSTS Y OW1 N EY UNA
UW1 N AA
UNICA UW1 N IY K AA
UNITI UW N IY1 T IY
UNO UW1 N OW
VACANZA V AA K AA1 N TS AA
VALIDA V AA1 L IY D AA
VANNO V AA1 NN OW
VANTAGGIO V AA N T AA1 JHJH OW
VECCHIA V EH1 KK Y AA
VEDERE V EY D EY1 R EY
VEDI V EY1 D IY
VEDIAMO V EY D Y AA1 M OW
VENDE V EY1 N D EY
VENT'ANNI V EY N T AA1 NN IY
VENTO V EH1 N T OW
VERDE V EY1 R D EY
VERSO V EH1 R S OW
VIALE V IY AA1 L EY
VICINA V IY CH IY1 N AA
VISTA V IY1 S T AA
VIVI V IY1 V IY
VOCE V OW1 CH EY
VOGLIONO V AO1 LHLH OW N OW
VOLENTIERI V OW L EY N T Y EH1 R IY
VOLESSI V OW L EY1 SS IY
VOLTATI V OW1 L T AA T IY
VOTI V OW1 T IY
VUOLE V W AO1 L EY
YOGA Y AO1 G AA
ZIA DZ IY1 AA
ZITTO TS IY1 TT OW
ZOO DZ AO1 OW